\documentclass[aps,pra,twocolumn,showpacs,showkeys]{revtex4-1}

\usepackage{graphicx}
\usepackage{amsmath}
\usepackage{amsfonts}
\usepackage{braket}
\usepackage{multirow}
\usepackage{float}
\usepackage{amsthm}
\begin{document}

\title[Nonlinear Optics in 2D Materials]{Third-order Optical Nonlinearity in\\ Two-dimensional Transition Metal Dichalcogenides}
\author{Sina Khorasani}
\affiliation{School of Electrical Engineering, Sharif University of Technology, Tehran, Iran\\
\'{E}cole Polytechnique F\'{e}d\'{e}rale de Lausanne, CH-1015, Lausanne, Switzerland}
\email{sina.khorasani@ieee.org}
\altaffiliation{Present address: Vienna Center for Quantum Science and Technology, University of Vienna, 1090 Vienna, Austria}

\begin{abstract}
We present a detailed calculation of the linear and nonlinear optical response of four types of monolayer two-dimensional (2D) transition-metal dichalcogenides (TMDCs), having the formula $\textrm{MX}_2$ with M=Mo,W and X=S,Se. The calculations are based on 6-band tight-binding model of TMDCs, and then performing a semi-classical perturbation analysis of response functions. We numerically calculate the linear $\chi_{\mu\nu}^{(1)}(-\omega;\omega)$ and nonlinear surface susceptibility tensors $\chi_{\mu\nu\zeta\eta}^{(3)} (-\omega_\Sigma;\omega_r,\omega_s,\omega_t)$ with $\omega_\Sigma=\omega_r+\omega_s+\omega_t$. Both non-degenerate and degenerate cases are studied for third-harmonic generation and nonlinear refractive index, respectively. Computational results obtained \textit{with no external fitting parameters} are discussed regarding two recent reported experiments on ${\rm MoS}_2$, and thus we can confirm the extraordinarily strong optical nonlinearity of TMDCs. As a possible application, we demonstrate generation of a $\frac{\pi}{4}-$rotated squeezed state by means of nonlinear response of TMDCs, in a silica micro-disk resonator covered with the 2D material. Our proposed method will enable accurate calculations of nonlinear optical response, such as four-wave mixing and high-harmonic generation in 2D materials and their heterostructures, thus enabling study of novel functionalities of 2D photonic integrated circuits. 
\end{abstract}

\pacs{42.65.-k; 71.20.Mq; 71.15.-m; 42.50.-p; 42.50.Dv}
\keywords{2D materials; Nonlinear optics; Transition metal dichalcogenides; Quantum optics}
\maketitle

\section{Introduction}

Following the discovery of the celebrated material, graphene in 2004, the field of two-dimensional (2D) materials has been rapidly expanding over the recent years \cite{0a,0b,0c,0d,0e,0f,0g,Mueller1,2DRev1}. Despite being only a monolayer thick, 2D materials exhibit extraordinary electronic, mechanical, thermal, spintronic, and in particular, optical properties. These offer novel and unprecedented applications which were not foreseen earlier such as new nonlinear capacitors \cite{Quantum} for using in cryogenic parametric amplifiers, circulators, and mixers, as well as a new type of capacitive qubit referred to as \textit{cubit} \cite{CUBIT} in quantum nonlinear superconducting circuits. Of primary importance, is the nonlinear optics of the 2D materials and structures made out of them, which is a matter of current deep investigations.

A number of researches discuss the second-order nonlinear susceptibility as well as second-harmonic generation in various 2D materials \cite{SHG1,SHG2,SHG3,SHG4,SHG5}. Similarly, the third-order nonlinear optical properties of graphene has been studied in many works \cite{0h,0i,0j,0k,0l,0m,0g1,0g2,0g3} and the two-dimensional nonlinear sheet susceptibility $\chi^{(3){\rm 2D}}$ or the conductivity $\sigma^{(3){\rm 2D}}$ tensor of graphene have been calculated. However, recent experiments \cite{0n} still disagree with the expected numerical estimates from theory within an order of magnitude. The nonlinearity of graphene is remarkably large, however, it needs to be gated to adjust its Fermi level to the resonance conditions. That implies for ungated structures, there is still a tendency to examine other 2D  materials as well. Similar discrepancies between measured sheet optical nonlinearity of alternative 2D materials and theoretical estimates is noticed by other researchers as well.

In this work, we present a detailed study of the nonlinear optical properties of 2D transition metal dichalcogenides (TMDCs) \cite{Manzeli}, with the general formula ${\rm MX}_2$ with M being a transition metal, here being either Molybdenum Mo or Tungsten W, and X being a chalcogen such as Sulphur S or Selenium Se. Based on a six-band Tight-Binding Hamiltonian, we obtain the nonlinear sheet susceptibility $\chi^{(3){\rm 2D}}$ of the TMDCs through semi-classical perturbation approach. 

The semi-classical method retains its validity for low illumination intensities, where Zener tunneling and semimetal transitions could be well ignored and dismissed from the carrier dynamics \cite{Tamya1,Tamya2,Tamya3}, leaving only multiphoton processes as important. This approximation also is limited by the finite span of optical wavelength and non-zero extent of atomic bonds, that is, the medium has to be effectively treated as continuous and its microscopic granular structure shall be ignored. For ultrashort wavelength optical excitations beyond the deep ultraviolet spectrum, this treatment is no longer valid.

Such nonlinear interactions \cite{Urb3} are of primary importance in study of valley-spin dynamics \cite{Urb1} and Kerr spectroscopy \cite{Urb2} of TMDCs. In all these last works, the TMDC monolayer has been sandwiched in protecting 2D Boron Nitride shields, which has greatly contributed to the visibility and sharpness of emission spectra \cite{Urb1,Urb2,Urb3,Urb4}.

The reported data on $\chi^{(3){\rm 2D}}$ of ${\rm MX}_2$ are very scattered up to four orders of magnitude, as it is shown to be strongly dependent on the growth method and the post-treatment process. This makes a conclusive evaluation out of reach at present. However, we notice agreement roughly within an order of magnitude between the computed numerical figure and the stronger one of the experimental values. 

We study the nonlinear optical properties of an ultralow-loss amorphous silica microdisk covered by ${\rm WSe}_2$, and define an effective nonlinear susceptibility $\chi_{\rm eff}^{(3)}$ for the structure. We show through detailed COMSOL calculations that the effective nonlinearity marked by $\chi_{\rm eff}^{(3)}$ is improved up to two orders of magnitude by placing the TMDC on top of the silica microdisk. This platform has been shown to be useful for experimental study of optical emission and excitonic properties of TMDCs \cite{Ex1}.

As an application example, we consider the squeezing property of light \cite{Hil} through four-wave mixing under unpumped and pumped configurations and discuss these scenarios. While calculations reveal the dominance of two-photon absorption over nonlinear Kerr self-phase modulation in 2D TMDCs, we show through detailed theory that an unorthodox $\frac{\pi}{4}-$rotated squeezed state of light is produced with an elongated elliptical Wigner distribution and theoretically unlimited squeezing. Without consideration of this $\frac{\pi}{4}-$rotation, both quadratures appear to be desqueezed. However, unlimited squeezing is possible only along one $\frac{\pi}{4}-$rotated quadrature and the orthogonal quadrature will always be strongly desqueezed.

\section{Theory \& Results}

\subsection{Band Structure}

In order to compute the linear and nonlinear susceptibility from first principles, we would need to have accurate knowledge of the electronic transitions, valleys, and spin-orbit interactions. The correct way to tackle this problem is to have an efficient code to derive the electronic band structure of the material, and since this information is going to be called upon quite frequently inside integration and summation loops, the computation has to be both accurate and very efficient. For this purpose, tight-binding (TB) scheme is very appealing since with correct implementation it could meet both criteria. 

The method TB is quite popular for low-dimensional carbon structures such as graphene and carbon nanotubes \cite{3a,3a1}, and six-band TB has been used for studying the band structure of hydrogenated graphane \cite{3b,3c}. In recent years, two-band \cite{3c2a,3c2b,2band} expansion based on $\textbf{k}\cdot\textbf{p}$ perturbation and L\"{o}wdin partitioning of Hamiltonian \cite{3w}, three-band \cite{3c3}, four-band \cite{3c4}, six-band \cite{3c6a,3c6b,3c6c,3c6d}, seven-band \cite{3c2b}, and eventually eleven-band \cite{3c11} TB models have been developed to calculate the band structure of TMDCs. However, six-band TB based \cite{3c6d} on Slater-Koster expressions \cite{3z} is ultimately chosen since it is a fairly accurate low-energy approximation to the extensive density functional theory (DFT) calculations, and thus expected to be both sufficiently accurate and efficient for our purpose. Moreover, the 6-band TB method, considers the effects of valley-dependent spin-orbit interaction of 2D TMDCs \cite{4}, or better known as trigonal warping \cite{3c2a,3c2b}. This method employs an orthonormal basis, and therefore does need inversion of the overlap matrix \cite{3a,3b} which eases out coding and efficiency at the same time. Since the details of this scheme is rather comprehensive, the reader is referred to the existing literature for further information \cite{3c6a,3c6b,3c6c,3c6d}. However, in Appendix {\ref{AppA}} we present the necessary ingredients briefly. The orthonormal bases here are
\begin{equation}
\label{eq1}
\ket{\bf \Psi}=\begin{Bmatrix}
d_{3z^2-r^2} \\
d_{x^2-y^2}\\
d_{xy}\\
p_{x}^+\\
p_{y}^+\\
p_{z}^-
\end{Bmatrix},
\end{equation}
\noindent
where $p_{l}^\eta=(p_l^t+\eta p_l^b)/\sqrt{2}$ with $t$ and $b$ referring to the top and bottom chalcogen atoms and $\eta=\pm$. For the sake of convenience, we rewrite the basis as $\ket{\bf \Psi}=\{\ket{\psi_n}\}, n=1\dots 6$. Obviously, the basis kets $\ket{\psi_n}$ are dependent on the 2D $\textbf{k}-$vector and spin $\sigma$, corresponding to the eigenvalues $E_n^\sigma(\textbf{k})$ which form the energy band structure. Hence, we have the relationship
\begin{equation}
\label{eq2} \braket{\psi_n^\sigma(\textbf{k})|\psi_m^{\sigma'}(\textbf{k})}=\delta_{nm}\delta_{\sigma\sigma'}.
\end{equation}

In general for honeycomb lattice with $C_{3v}$ symmetry such as graphene \cite{3a1} and in absence of chirality, it is normally sufficient to compute the band structure over the irreducible Brillouin zone, which is only $1/12$ of the first Brillouin zone. However, monolayer TMDCs are non-centrosymmetric and satisfy a different spatial group denoted as $D_{3h}^1$ \cite{4f0}, which causes spin-valley coupling. As a result, the irreducible zone is only $1/6$ of the first Brillouin zone.

\begin{figure}
	\centering
	\includegraphics[width=2.2in]{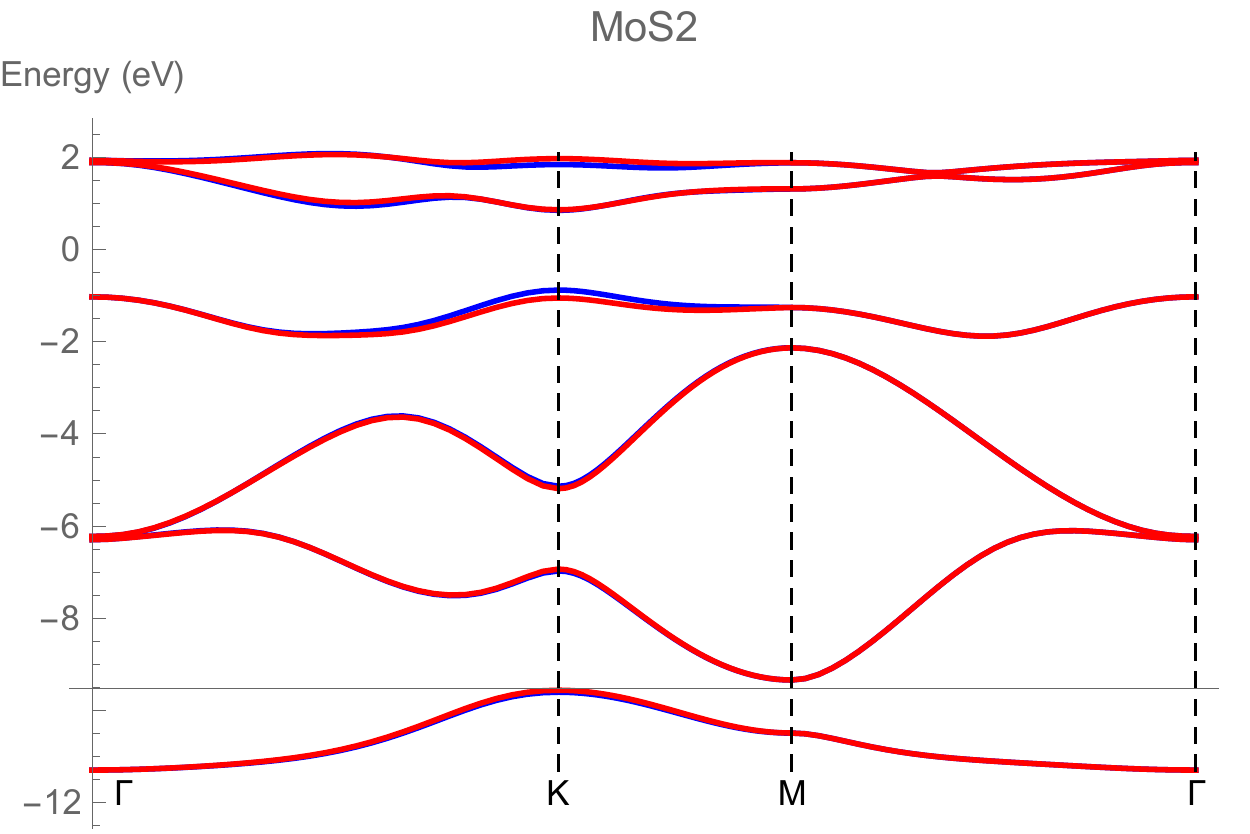}
	\includegraphics[width=2.2in]{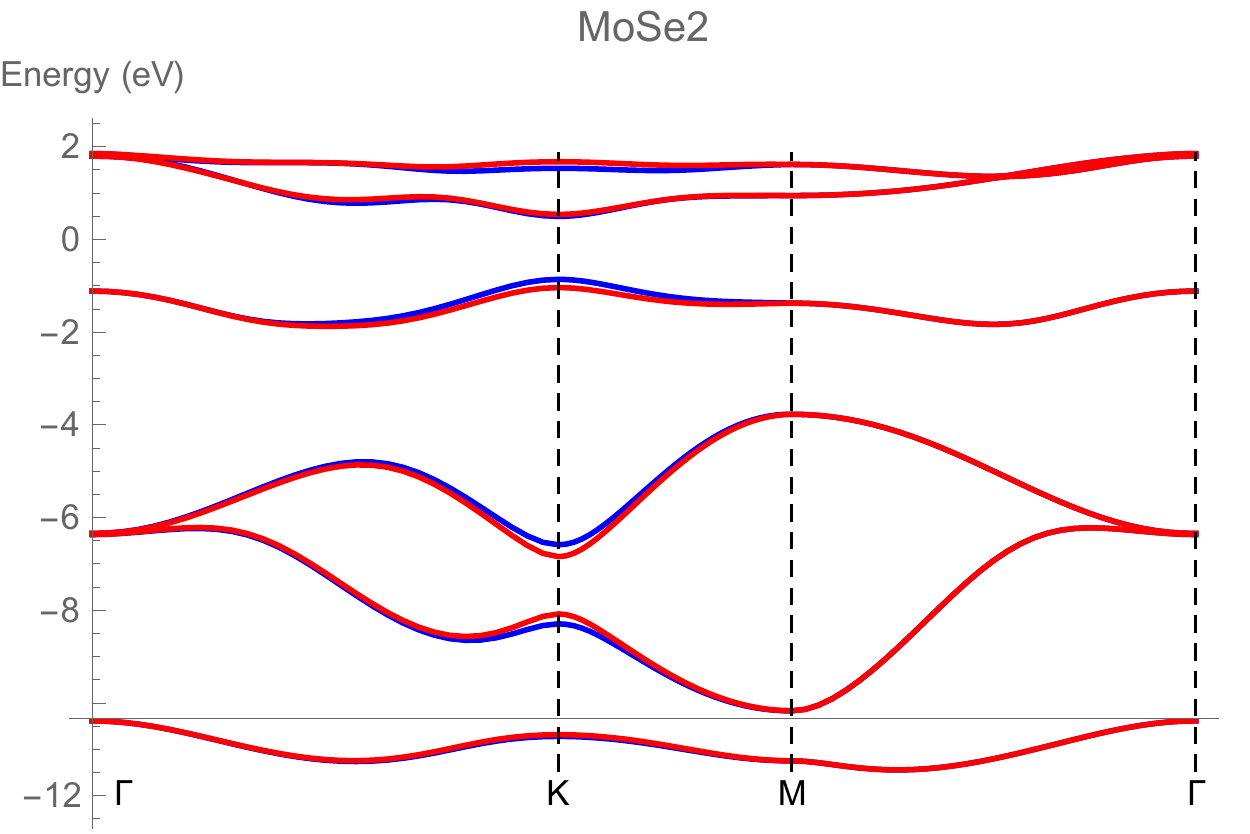}
	\includegraphics[width=2.2in]{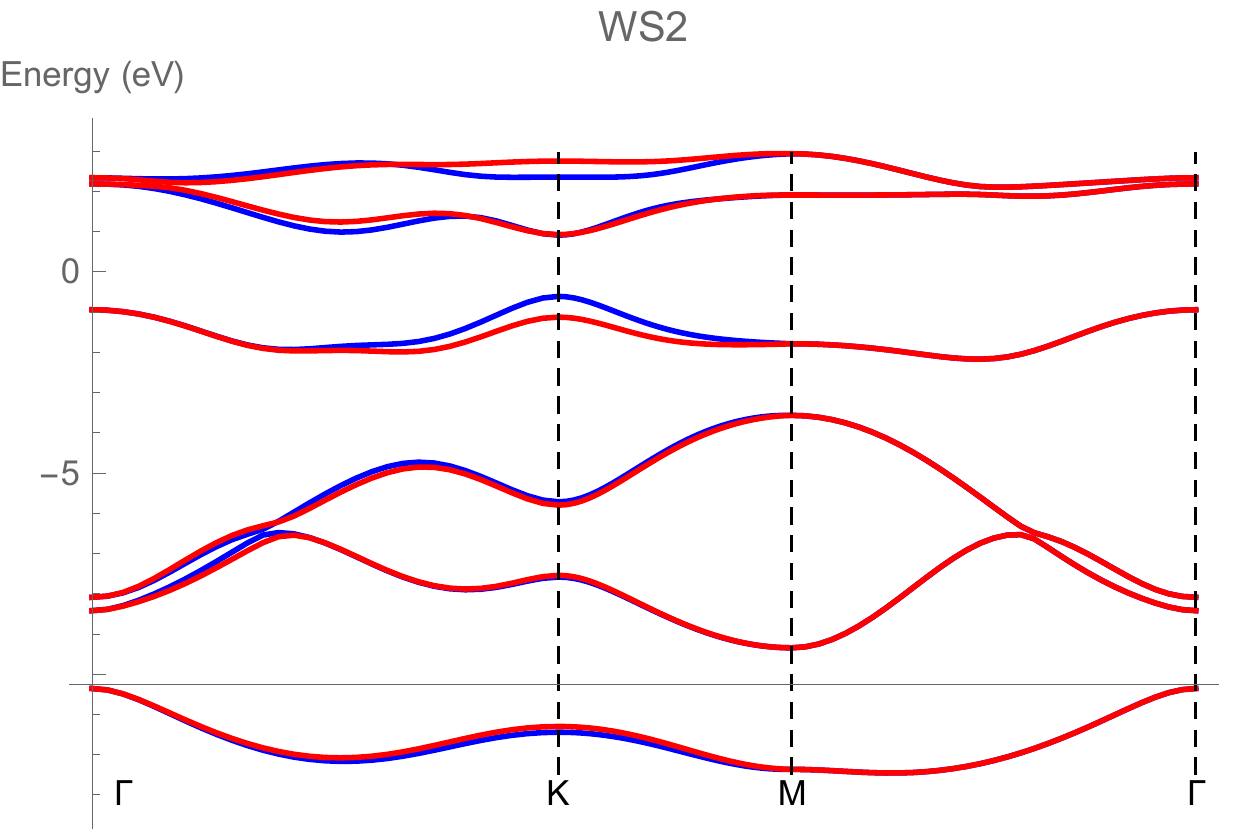}
	\includegraphics[width=2.2in]{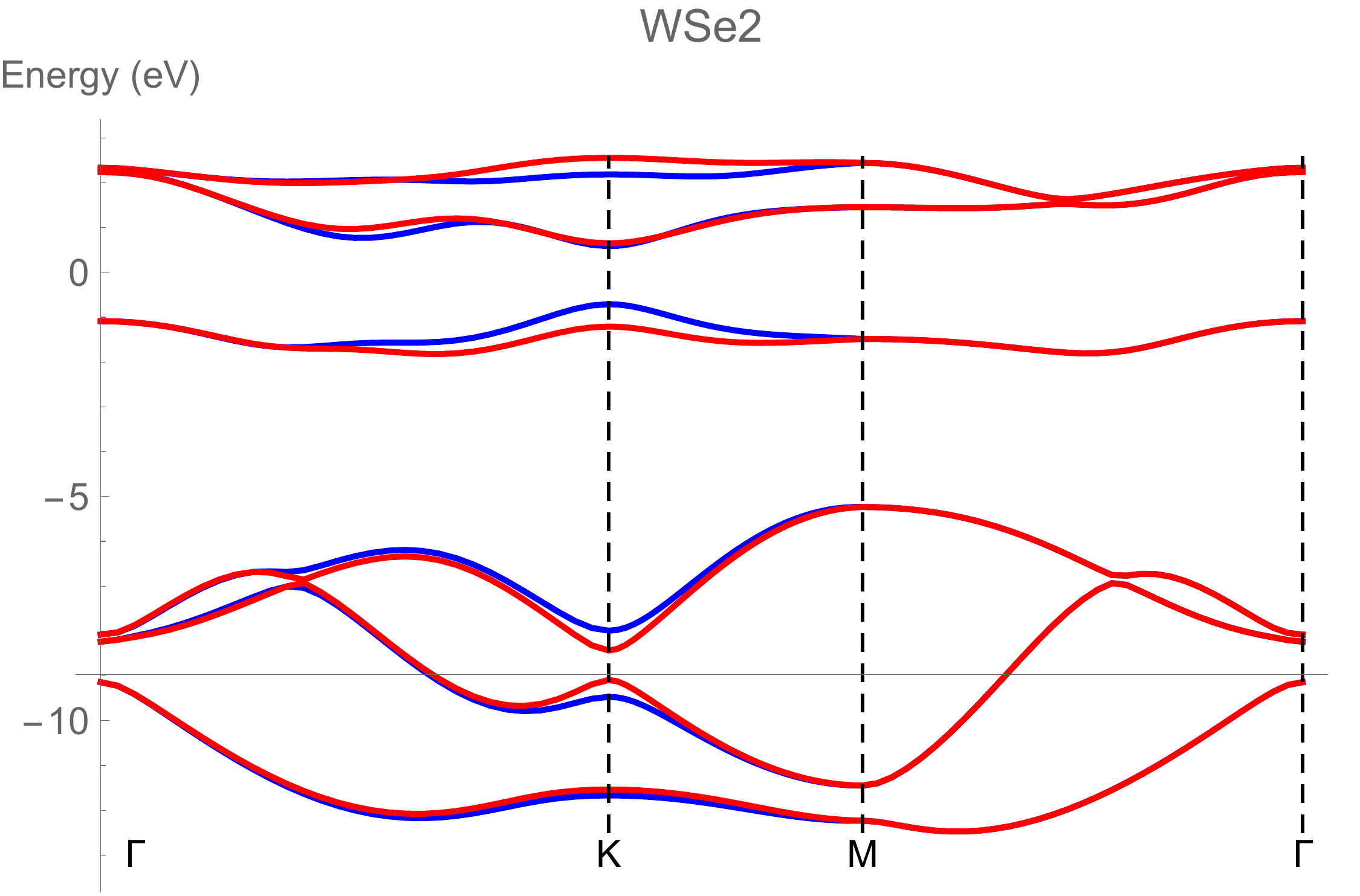}
	\caption{Band structures of four basic TMDCs, based on the 6-band TB model \cite{3c6d}. Red and blue lines correspond to the spin down and up bands.\label{Fig1}}
\end{figure}

\begin{figure}
	\centering
	\includegraphics[width=2.2in]{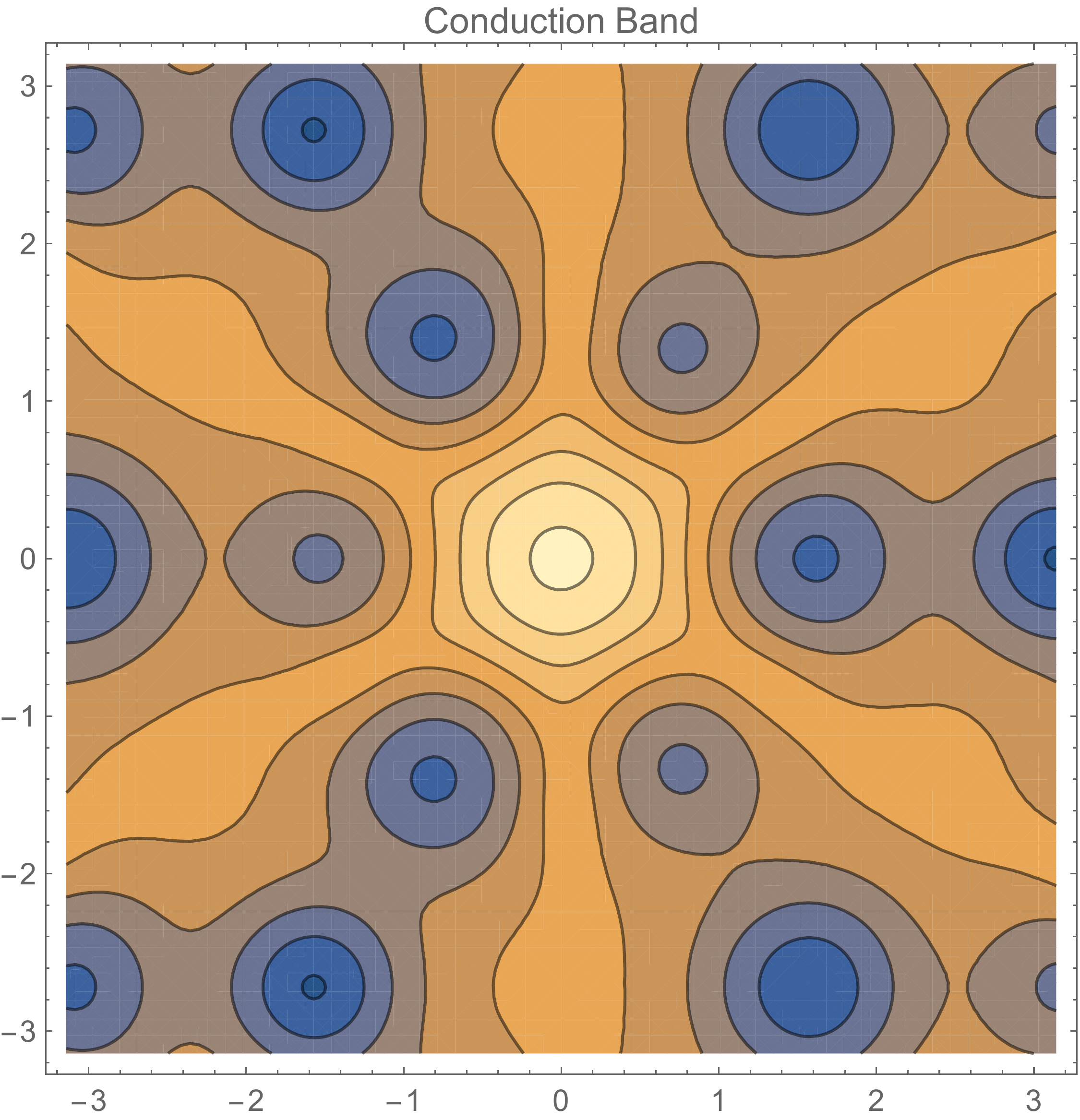}
	\includegraphics[width=2.2in]{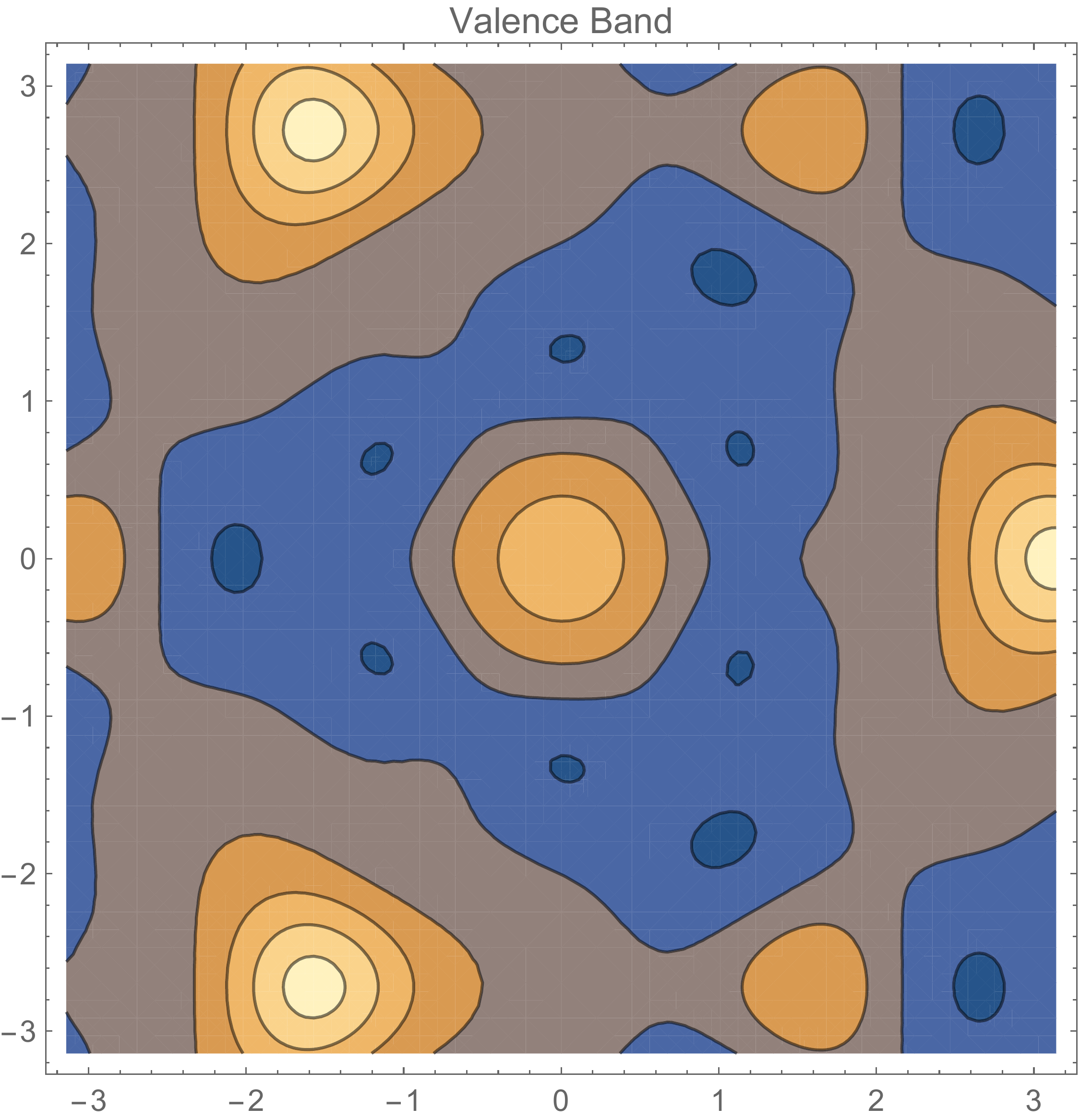}
	\caption{Illustration of valley-spin coupling in $\textrm{WSe}_2$ as an example, based on the 6-band TB model \cite{3c6d}. The effect of trigonal warping is seen to be strongly pronounced for the valence band, where  K and ${\rm K}'$ behave very differently.\label{Fig2}}
\end{figure}

The calculated band structures are shown in Fig. \ref{Fig1} for the four basic types of TMDCs $\textrm{MX}_2$ with M=Mo,W and X=S,Se. Here, red and blue curves correspond to respectively down and up spin polarizations. Interestingly, the calculations preserve valley-dependent spin-orbit interaction or the trigonal warping of TMDCs, which is strongly present in the valence bands at K and ${\rm K}'$. To illustrate this, we calculate the entire first Brillouin zone of $\textrm{WSe}_2$, in which spin-orbit interaction is highly different at K and ${\rm K}'$. As a result of this symmetry breaking, the band structure shows a $C_{3v}$ point-group symmetry, instead of the expected $C_{6v}$, and by changing the direction of spin the band structure rotates by $\pi/3$, thereby interchanging the roles of K and ${\rm K}'$. The conduction and valence bands of  $\textrm{WSe}_2$ are illustrated in Fig. \ref{Fig2}. 

\subsection{Linear Susceptibility}

The concept of an interface occupied by a dielectric having finite surface electric and magnetic susceptibilities has been originally investigated in a pair of papers published in 1994 \cite{4f1,4f2}. Independently and still a few years before celebrated discovery of the first 2D material, graphene, the optical properties, possible modulation and switching applications of such media, referred to as the \textit{conducting interfaces} were explored by a series of papers by the author \cite{4a,4b,4c,4d,4e,4f}. The transfer matrix method has been reformulated so far independently by many authors to tackle the problem of optical wave refraction from layered structures containing 2D materials \cite{4f2,4b,4g,4h,4i} and even more recently by Morano \cite{4j,4k} for measurement of the surface conductivity of graphene.

Based on the theory of conducting interfaces, an ultrathin 2D material could be treated by a discontinuity in tangential electromagnetic fields across the interface. However, the corresponding surface susceptibility or $\chi^\textrm{2D}$ with the dimension of meter, or equivalently, surface conductivity $\sigma_\textrm{s}=j\epsilon_0\omega\chi^\textrm{2D}$ with the dimension of $\Omega^{-1}$ should be known. 

Here, we are able to compute the surface susceptibility tensor elements $\chi^\textrm{2D}(\omega)=\chi^{(1)\rm 2D}(-\omega;\omega)$ of TMDCs using the expression
\begin{eqnarray}
\label{eq3}
\chi^{(1)\rm 2D}_{\mu\nu}&&(-\omega;\omega)=\frac{1}{\epsilon_0\omega^2 A_{\rm UC}}\times \\ \nonumber 
&&\sum_{m n\sigma}\frac{1}{A_{\rm BZ}}\iint_{\rm BZ}\frac{f_n^\sigma({\bf k})-f_m^\sigma({\bf k})}{E_{mn}^\sigma({\bf k})-\hbar\omega}j^{\mu\sigma}_{mn}({\bf k})j^{\nu\sigma}_{nm}({\bf k})d^2k,
\end{eqnarray}
\noindent
where $E_{mn}^\sigma({\bf k})=E_m^\sigma({\bf k})-E_n^\sigma({\bf k})$, $f_m^\sigma({\bf k})=\{1+\exp[E_m^\sigma({\bf k})-E_{\rm F}/k_{\rm B}T]\}^{-1}$ is the Fermi-Dirac distribution with $E_{\rm F}$ being the Fermi energy (here equal to zero for the intrinsic semiconductor), and $k_{\rm B}T$ being the thermal energy. The matrix elements $j^{\mu\sigma}_{mn}({\bf k})$ are related to the current operator $\hat{j}_{\mu}^{\sigma}({\bf k})$ as
\begin{eqnarray}
\label{eq4}
j^{\mu\sigma}_{mn}({\bf k})&=&\braket{\psi_m^\sigma({\bf k})|\hat{j}_{\mu}^{\sigma}({\bf k})|\psi_n^\sigma({\bf k})},\\ \nonumber
\hat{j}_{\mu}^{\sigma}({\bf k})&=&-\frac{q}{\hbar}\frac{\partial\mathbb{H({\bf k};\sigma)}}{\partial k_{\mu}},
\end{eqnarray}
\noindent
in which $k_{\mu}$ are elements of the vector ${\bf k}$. Moreover, $q$ is the electronic charge and $\mathbb{H}({\bf k};\sigma)$ is the spin-dependent Hamiltonian in ${\bf k}-$space, which in our TB formalism appears as a $6\times 6$ matrix. More on the details of this operators could be found in the literature \cite{eqm3}. In (\ref{eq3}), $\epsilon_0$ is the permittivity of vacuum, $A_{\rm UC}$ is the area of unit cell, and $A_{\rm BZ}=(2\pi)^2/A_{\rm UC}$ is the area of first Brillouin zone \cite{Avo,Avo1,Avo2}. That implies the expression $N_{\rm A} A_{\rm UC}$ is the area occupied by one mole of the 2D material, where $N_{\rm A}$ is Avogadro's constant \cite{Avox,Avo,Avo1,Avo1x,Avo2,Avo2x,Av1}. One can easily verify that $A_{\rm UC}=\sqrt{3}a^2/2$ where $a$ is the 2D crystal's lattice constant, being equal to the distance between the two nearest metal atoms.

In the third-order nonlinear processes, the location of Fermi-energy has very little observable effect on the Kerr coefficient, and much less on the third-harmonic generation coefficient. This is a fact observed through extensive numerical experiments. Only if the Fermi level is deeply into the conduction or valence bands, that is, the semiconductor is made degenerate, the result would become different. Normally, since chemical doping is not in practice for TMDCs, any shift in the Fermi level would be triggered by electrostatic gating. This method of charge depletion or accumulation, however, is typically unpractical for making a 2D TMDC degenerate. The author believes that there is no practical reason to be concerned about the effect of extrinsic Fermi level as opposed to the intrinsic case with $E_F=0$.

When the TMDC is not under strain, then linear susceptibility $\chi^{(1)\rm 2D}_{\mu\nu}(-\omega;\omega)=\chi^{(1)\rm 2D}(\omega)\delta_{\mu\nu}$ is a simple scalar and not a tensor quantity. This fact together with the spin-valley coupling of carriers can be used to simplify (\ref{eq3}) a bit as
\begin{eqnarray}
\label{eq5}
\chi^{(1)\rm 2D}&&(-\omega;\omega)=\frac{3}{\epsilon_0\omega^2 A_{\rm UC}}\times \\ \nonumber 
&&\sum_{m n\sigma\tau}\frac{1}{A_{\rm BZ}}\iint_{\rm IRBZ}\frac{f_n^{\sigma\tau}({\bf k})-f_m^{\sigma\tau}({\bf k})}{E_{mn}^{\sigma\tau}({\bf k})-\hbar\omega}|j^{x{\sigma\tau}}_{mn}({\bf k})|^2d^2k,
\end{eqnarray}
\noindent
Here, the integration on the reciprocal plane is taken only over the irreducible Brillouin zone, and summation on valley index $\tau=\pm$ is for selection between ${\rm K}$ and ${\rm K}'$ valleys. Since valley contributions from opposite spins are simply equal, one may take advantage of symmetry considerations and drop the summation on $\tau$, just multiplying the whole expression on the right by a factor of 2, thus arriving at
\begin{eqnarray}
\label{eq6}
\chi^{(1)\rm 2D}&&(-\omega;\omega)=\frac{6}{\epsilon_0\omega^2 A_{\rm UC}}\times \\ \nonumber 
&&\sum_{m n\sigma}\frac{1}{A_{\rm BZ}}\iint_{\rm IRBZ}\frac{f_n^{\sigma}({\bf k})-f_m^{\sigma}({\bf k})}{E_{mn}^{\sigma}({\bf k})-\hbar\omega}|j^{x{\sigma}}_{mn}({\bf k})|^2d^2k.
\end{eqnarray}

\begin{figure}[ht!]
	\centering
	\includegraphics[width=2.2in]{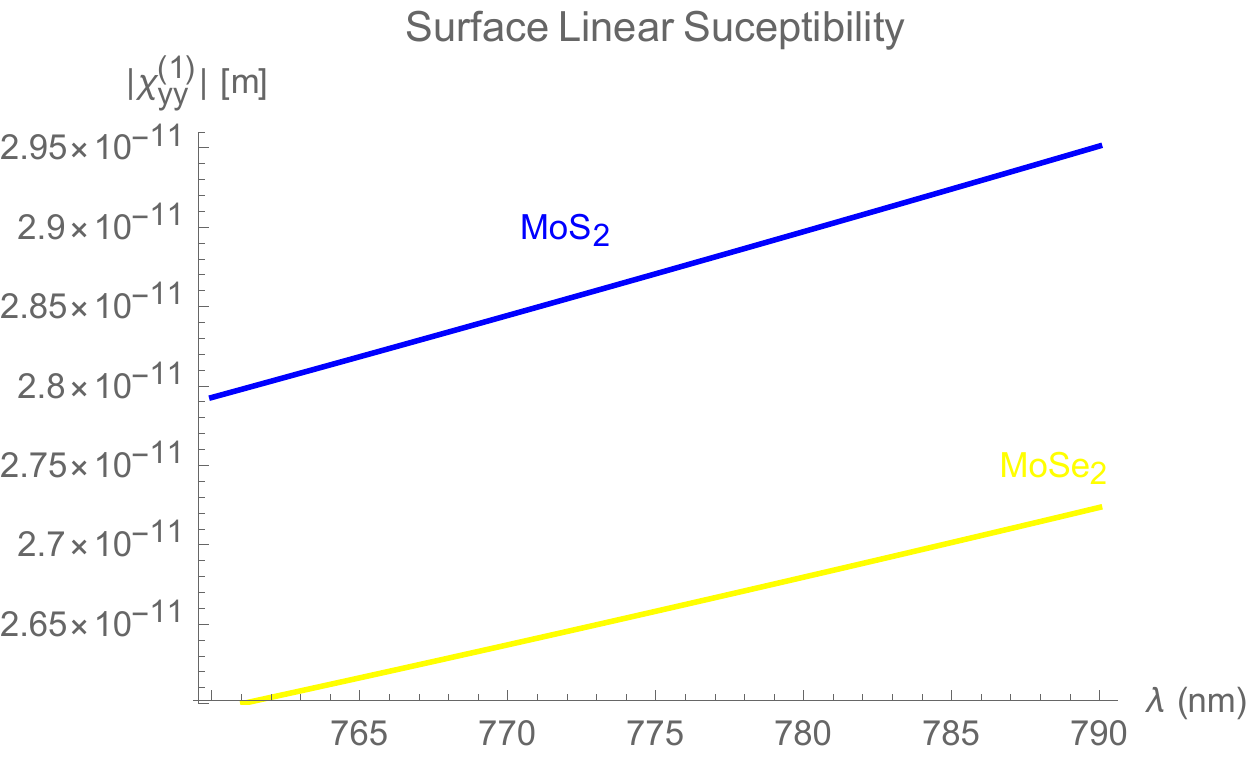}
	\includegraphics[width=2.2in]{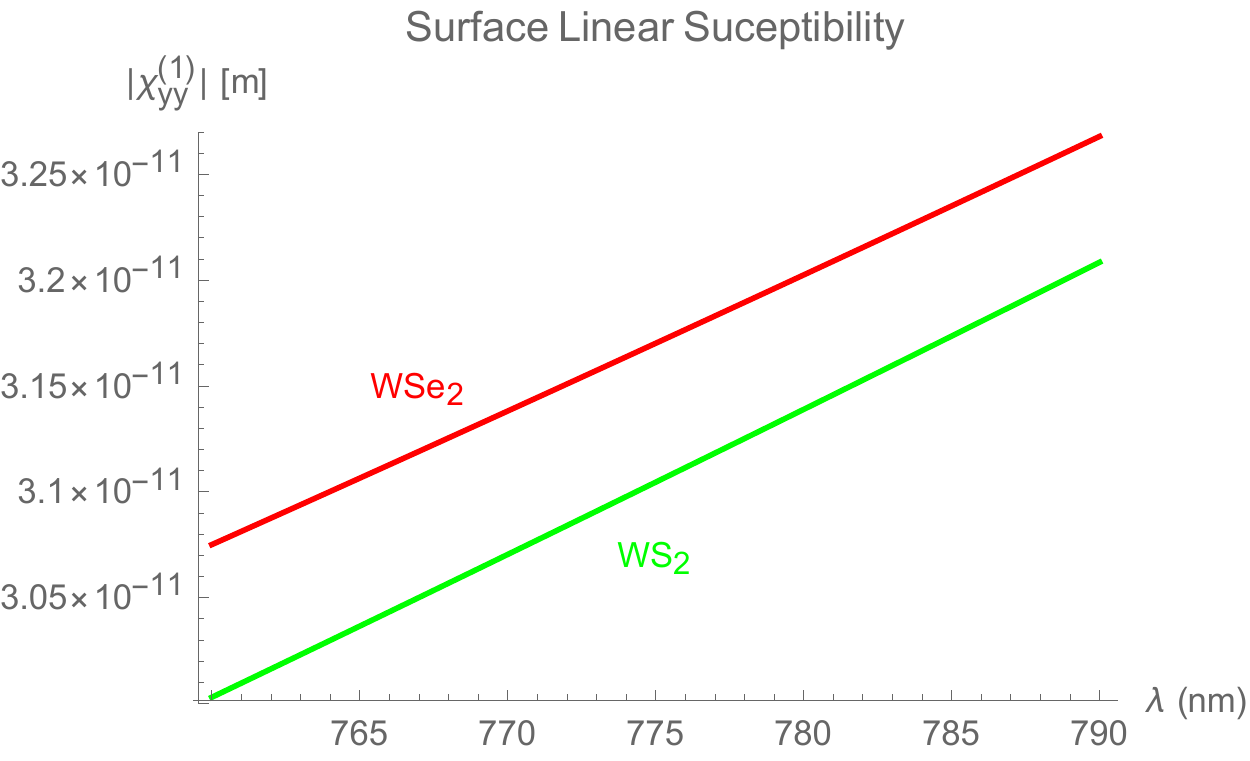}
	\caption{Typical values of first-order linear susceptibility $\chi^{(1)}(-\omega;\omega)$ for different TMDCs. Contribution to the absolute value is entirely from the real part.\label{Fig2Lin}}
\end{figure}

Results of computations for the linear surface susceptibilities of the four basic TMDCs in the wavelength range 760-790nm is illustrated in Fig. \ref{Fig2Lin}. Although the absolute values are here shown, imaginary parts of $\chi^{(1)\rm 2D}(-\omega;\omega)$ in the desired range is effectively zero.

\subsection{Nonlinear Susceptibility}

Folllowing the standard perturbation scheme \cite{6,7,7a,7b,7c,7d,7e,7f,7g,7h,7i,7j,7k}, we have developed a code in Mathematica to compute the non-degenerate four-wave mixing (FWM) third-order nonlinear susceptibility $\chi_{mnpq}^{(3)} (-\omega_\Sigma;\omega_r,\omega_s,\omega_t )$ where $\omega_\Sigma=\omega_r+\omega_s+\omega_t$ for any of the widely used TMDCs. When studying 3rd harmonic generation,  $\chi_{mnpq}^{(3)} (-3\omega;\omega,\omega,\omega)$ must be calculated, while for degenerate nonlinear propagation,  $\chi_{mnpq}^{(3)} (-\omega;\omega,-\omega,\omega)$ should be found. Expressions are given in the Appendix {\ref{AppB}}. It is worth mentioning that the method of equations of motions \cite{eqm1,eqm2,eqm3} could also be used to tackle the dynamics of nonlinear optics and four-wave mixing in semiconducting materials, however, the formalism does not directly give in an expression for the nonlinear susceptibility.

\begin{figure}
	\centering
	\includegraphics[width=2.2in]{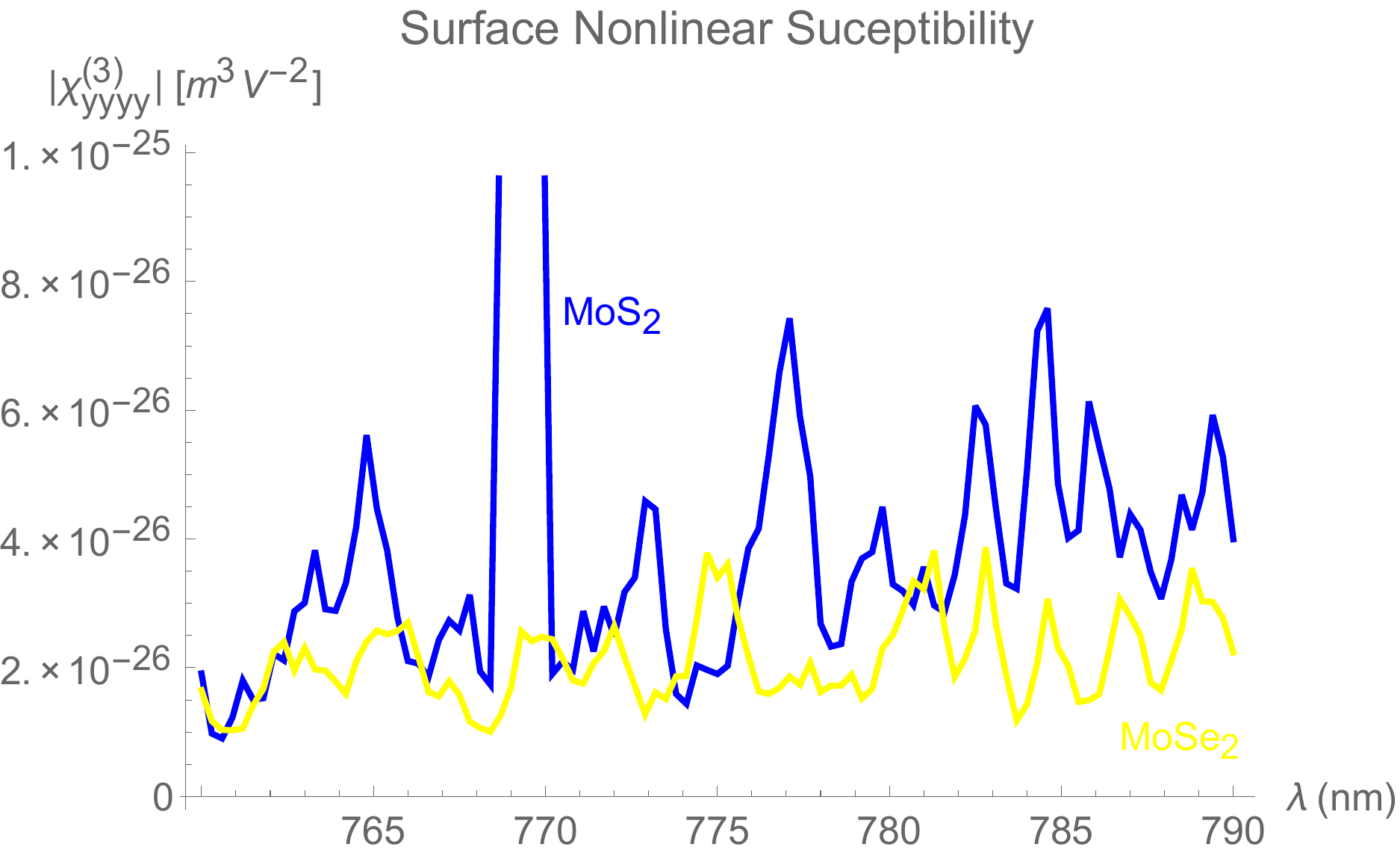}
	\includegraphics[width=2.2in]{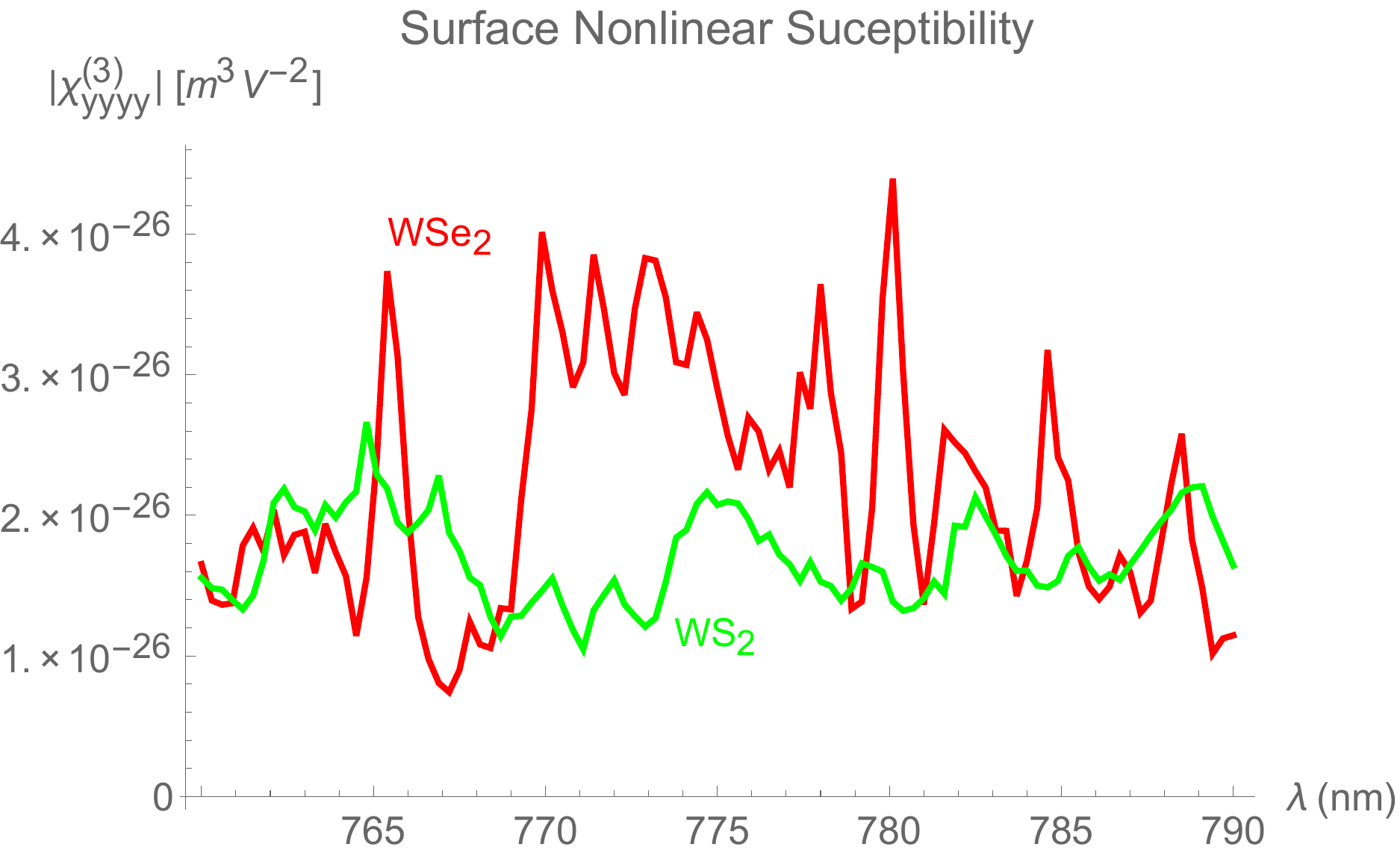}
	\caption{Typical values of third-order nonlinear susceptibility $\chi^{(3)}(-3\omega;\omega,\omega,\omega)$ for different TMDCs. Contribution to the absolute value is entirely from the real part.\label{Fig3}}
\end{figure}

Figure \ref{Fig3} represents typical calculated values from semi-classical perturbation theory of nonlinear interactions. Similarly, Fig. \ref{Fig4} presents the typical nonlinear coefficient in the telecommunication window around the wavelength 1550nm. The uniform convergence of the code has been demonstrated in Fig. \ref{Fig5} versus computational grid. These set of figures present absolute value of the tensor component $|\chi^{(3)}_{yyyy}(-3\omega;\omega,\omega,\omega)|$ corresponding to the third-harmonic generation. Calculation of other non-zero tensor components has also been done, but nor presented for the sake of brevity. Further discussions on the relationship among non-zero tensor components is given in the following.

\begin{figure}
	\centering
	\includegraphics[width=2.2in]{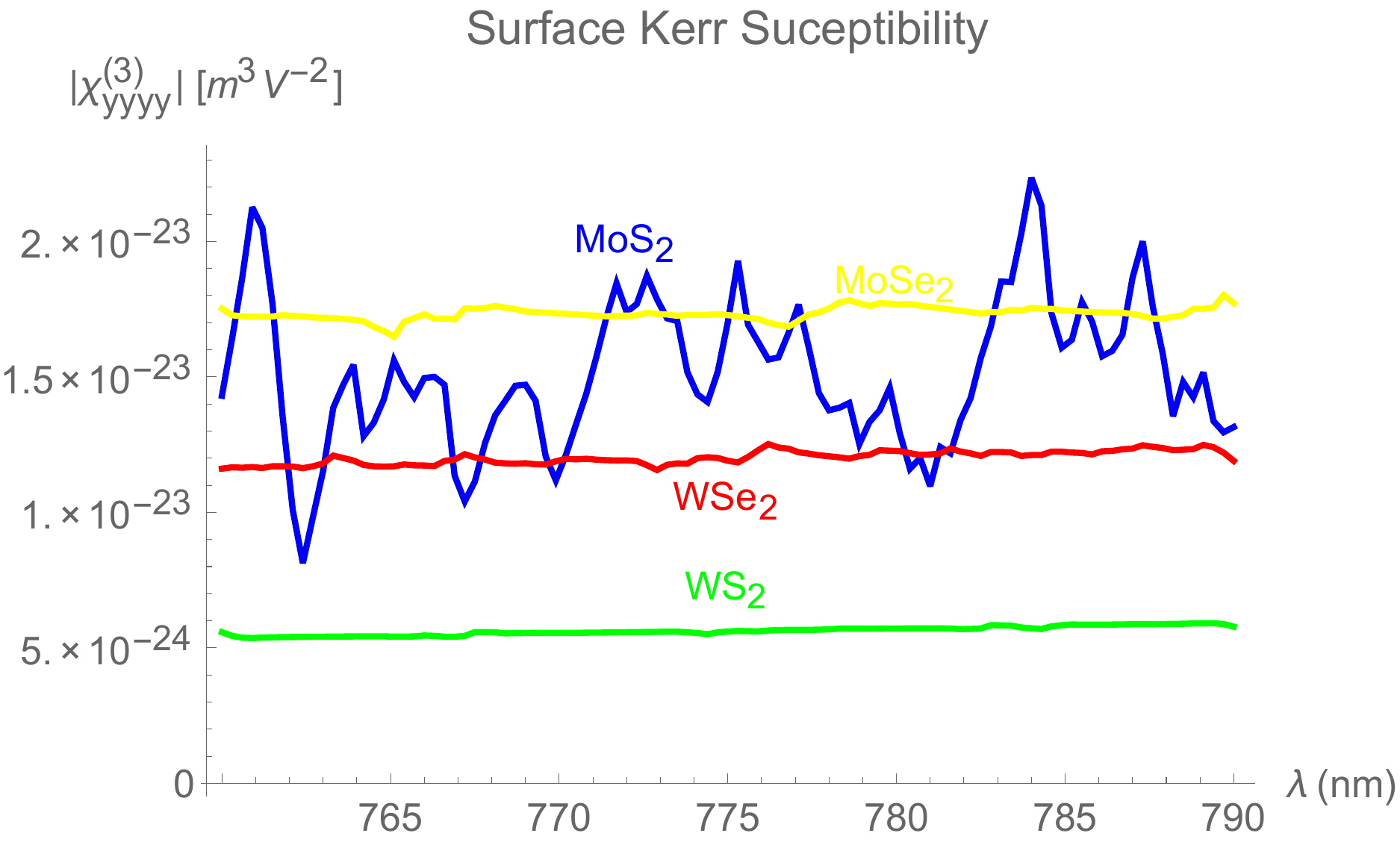}
	\caption{Typical values of Kerr nonlinear susceptibility $\chi^{(3)}(-\omega;\omega,-\omega,\omega)$ for different TMDCs. Contribution to the absolute value is entirely from the imaginary part.\label{Fig3Kerr}}
\end{figure} 

\begin{figure}
	\centering
	\includegraphics[width=2.2in]{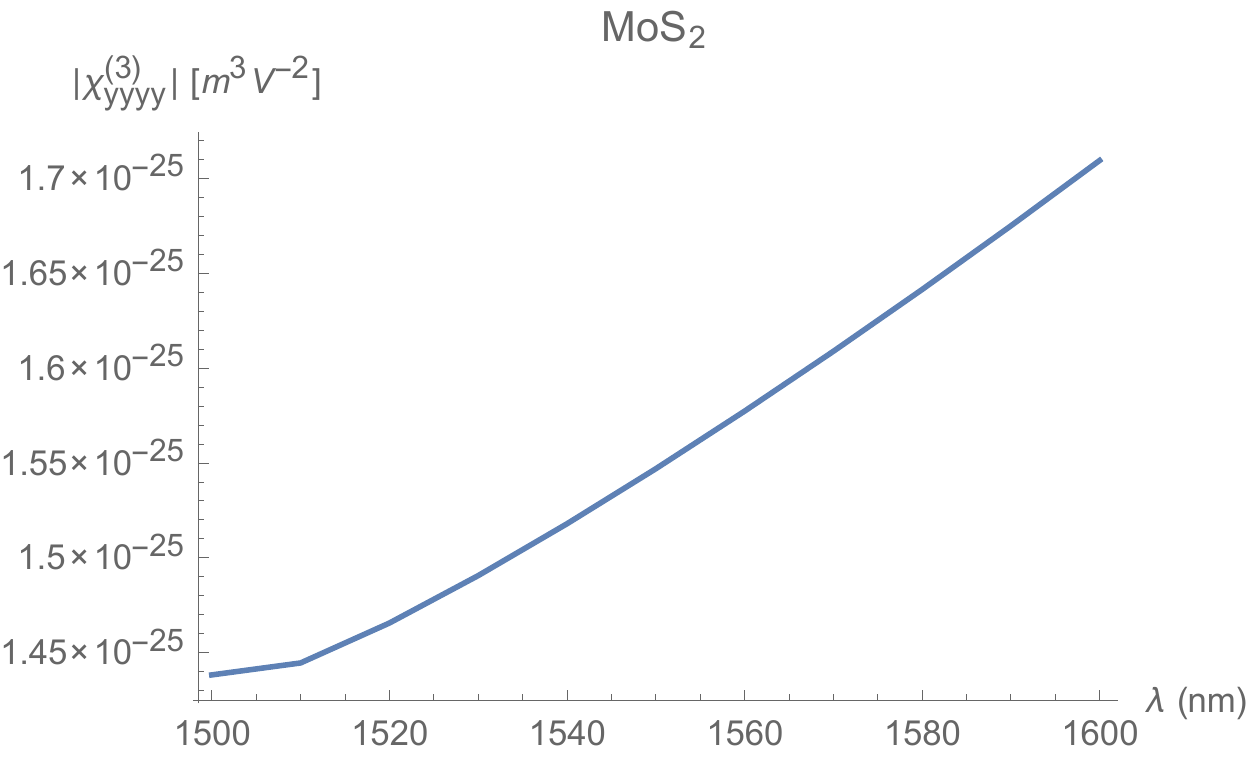}
	\caption{The calculated third-order nonlinear susceptibility $\chi^{(3)}(-3\omega;\omega,\omega,\omega)$ for ${\rm MoS}_2$ in the infrared telecommunication window. Contribution to the absolute value is entirely from the real part.\label{Fig4}}
\end{figure} 

\begin{figure}
	\centering
	\includegraphics[width=2.2in]{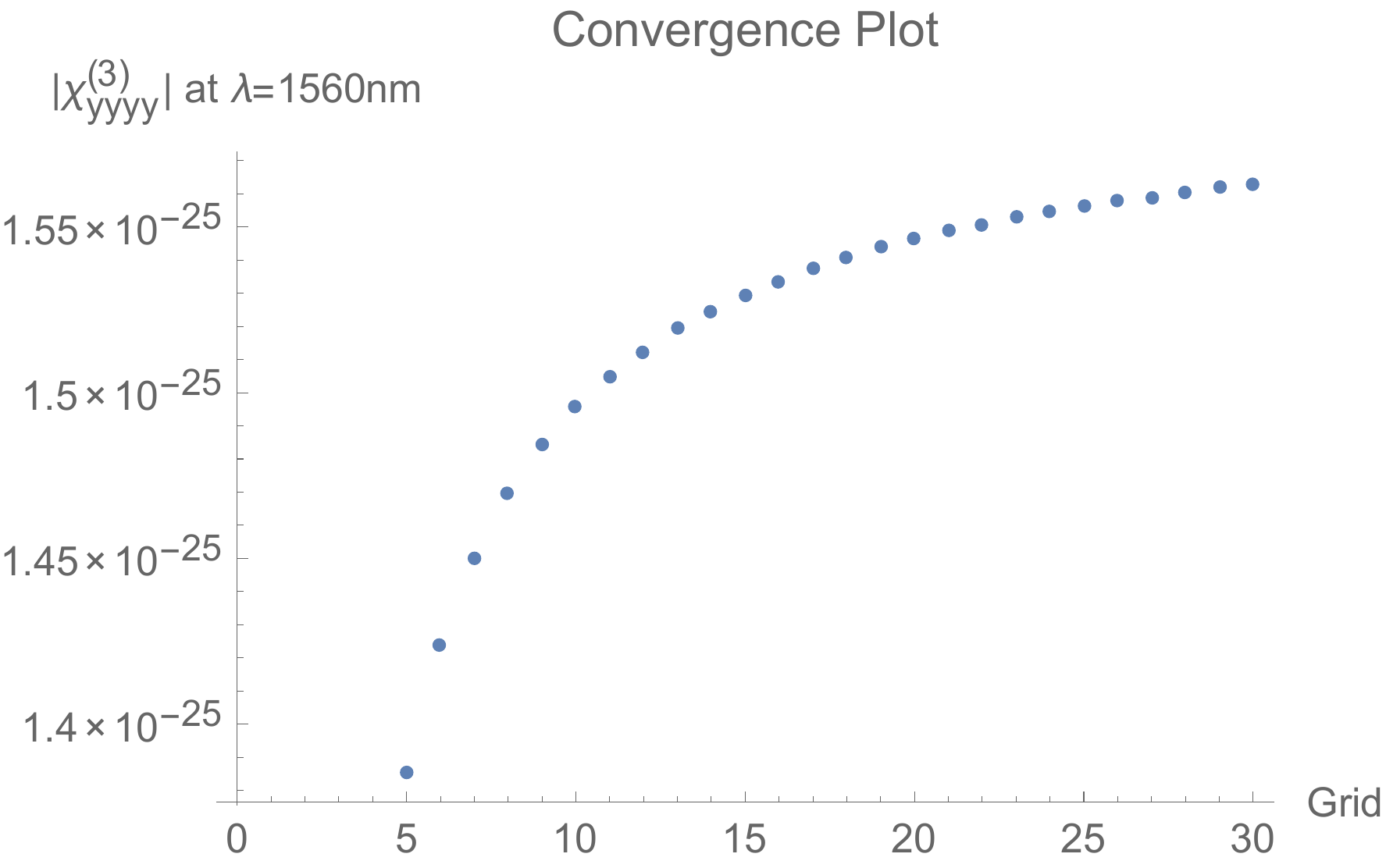}
	\caption{Typical convergence of the code at the wavelength of 1560nm versus resolution of the computational grid, corresponding to Fig. \ref{Fig4}.\label{Fig5}}
\end{figure} 

Resonances in $\chi_{mnpq}^{(3)} (-3\omega;\omega,\omega,\omega)$ as shown in Fig. 2, correlate well with the resonances in linear surface susceptibility $\chi^{(1){\rm 2D}}(-\frac{1}{n}\omega;\frac{1}{n}\omega)$ with $n=1,2,3$, as well as the possible such direct transitions across the band structure. For the case of degenerate nonlinear coefficient $\chi_{mnpq}^{(3)} (-\omega;\omega,-\omega,\omega)$, resonances are possible only at $n=1,2$, and were found to be much less pronounced. The code is able to compute individual tensor components as arbitrated such as $\chi_{ijkl}^{(3){\rm 2D}}$.

Computations on the Kerr nonlinear susceptibility $\chi^{(3)}_{mnpq}(-\omega;\omega,-\omega,\omega)$ reveals that it is entirely imaginary in the wavelength range of interest, that is, it is the two-photon absorption which is the dominant Kerr nonlinear effect. Furthermore, this coefficient turns out to be a scalar like $\chi^{(3)}(-\omega;\omega,-\omega,\omega)$, as is discussed later below. Typical values for different types of TMDCs are illustrated in Fig. \ref{Fig3Kerr}. Comparing to Fig. \ref{Fig3}, much less oscillations are seen, and that is because of the much less number of available resonances with transitions among band edges. For the case of ${\rm MoS}_2$, in particular, many resonances occur as shown in Fig. \ref{Fig3Kerr}, which can be attributed to the bandgap of this material being the smallest of the four studied TMDCs. As a result, for a given photon energy above the energy gap, much more possible triple transitions required for a third-order nonlinear process become available. Evidently, selection of a photon energy sufficiently small, that is the case in Fig. \ref{Fig4}, no such behavior is seen, simply because of the absence in availability of corresponding possible transitions. 

Furthermore, the magnitude of Kerr nonlinear index is significantly larger than the third-harmonic generation. It is thus to be noticed well that these two parameters of Kerr nonlinearity and third-harmonic generation, although related, but are essentially much different in both their physics and behavior. Actually, these two values could be quite far apart in phase and many orders in magnitude.

We here point out again that the Kerr nonlinear susceptibility $\chi^{(3)}(-\omega;\omega,-\omega,\omega)$ and Third-harmonic generation susceptibility $\chi^{(3)} (-3\omega;\omega,\omega,\omega)$ are not the same. While both are having the third-order, these correspond to entirely different processes. This difference can be understood as follows. 

The Kerr nonlinear susceptibility only describes a process in which the local permittivity of the medium is linearly dependent on the local intensity of light, such as $\epsilon(\textbf{E})=n^2\epsilon_0+|{\bf \bar{E}}|^2\chi^{(3)}$, where ${\bf \bar{E}}$ is the phasor of the electromagnetic field. Hence, this effect can for instance cause local increase or decrease of the effective refractive index depending on the sign of $\Re[\chi^{(3)}]$, which correspondingly lead to self-focusing or self-defocusing phenomena. While higher harmonics inevitably may weakly arise under such conditions, because of the inherent nonlinearity in the propagation, it is only the first harmonic which undergoes self-field propagation issues. More accurately, ${\bf \bar{E}}^{(3)}(\textbf{r})e^{i\omega t}=[{\bf \bar{E}}^{(1)}(\textbf{r})e^{i\omega t}:\chi^{(3)}:{\bf \bar{E}}^{(1)\ast}(\textbf{r})e^{-i\omega t}]{\bf \bar{E}}^{(1)}(\textbf{r})e^{i\omega t}$.

However, the third-order nonlinear susceptibility directly is connected to the local amplitude of the harmonic, as a result of locally present electromagnetic field. Hence, one could imagine that the third-harmonic be related to the first harmonic as ${\bf \bar{E}}^{(3)}(\textbf{r})e^{i3\omega t}=[{\bf \bar{E}}^{(1)}(\textbf{r})e^{i\omega t}:\chi^{(3)}:{\bf \bar{E}}^{(1)}(\textbf{r})e^{i\omega t}]{\bf \bar{E}}^{(1)}(\textbf{r})e^{i\omega t}$.

Since the physical phenomena behind these two mechanism are not identical, one would naturally expect that $\chi^{(3)}(-\omega;\omega,-\omega,\omega)\neq\chi^{(3)} (-3\omega;\omega,\omega,\omega)$. Sometimes, this fact is not sufficiently made clear in the literature. 

\subsubsection{Critical Field}

Defining a critical electric field as $E_c^2=|\chi^{(1)}/\chi^{(3)}_{xxxx}|$ as a measure of required light’s electrical field to reach the onset of third-order nonlinearity, we observe that for the TMDCs this value falls within the typical range of 7-8kV/m, regardless of the choice of the particular material. For all the studied cases in the wavelength range of 750-790nm, the approximations $\Im[\chi_{mn}^{(1)}]=0$,  $\Im[\chi_{mnpq}^{(3)}](-3\omega;\omega,\omega,\omega)=0$, as well as $\Re[\chi_{mnpq}^{(3)}](-\omega;\omega,-\omega,\omega)=0$ apply well to the linear and nonlinear surface polarizabilities. We use the conducting interface formulation \cite{4a,4b} to carry out any electromagnetic analysis of monolayer 2D materials.

\subsubsection{Analogy with Bulk Values}

In this wavelength range of interest and for the linear response, the typical strength of linear susceptibility $\chi^{(1)}$ is within the range of $3.5-4.4\times 10^{-8} {\rm m}$, so that having an effective monolayer thickness of $t_{\rm 2D}$ we may assign an effective refractive index of $n=\sqrt{1+[\chi^{(1)}/t_{\rm 2D}]}$ to the ultrathin TMDC layer. Since, $t_{\rm 2D}$ is only of the order of typically $7-10$\AA, then effective refractive index is expected to be about only $n\approx 6-8$ in the wavelength range presented here. However, we notice that at longer infrared wavelengths, this value dramatically increases to higher values. 

Referring to Fig. \ref{Fig3Kerr}, TMDCs have an extremely large surface nonlinearity $\chi^{(3){\rm 2D}}$ in the range of $5\times 10^{-24} {\rm m^3 V^{-2}}$ to $2\times 10^{-23} {\rm m^3 V^{-2}}$ \cite{5,6}, which when divided by the typical thickness of a monolayer of the order of 0.5nm, give rise to an effective nonlinearity of the order of $10^{-14} {\rm m^2 V^{-2}}$ to $10^{-13} {\rm m^2 V^{-2}}$. Given that this value is resulting from pure electronic polarization, this is quite remarkably large when put into perspective of expected values of the order of $10^{-22} {\rm m^2 V^{-2}}$ \cite{7}. 

This places the expected nonlinearity of 2D TMDCs, to many orders of magnitude stronger the range of III-V semiconductors such as GaAs $1.4\times 10^{-22} {\rm m^2 V^{-2}}$, and than that of Diamond $2.5\times 10^{-21} {\rm m^2 V^{-2}}$, fused silica $2.5\times 10^{-22} {\rm m^2 V^{-2}}$, and even GaP at 577nm $2.93\times 10^{-18}{\rm m^2 V^{-2}}$ \cite{7c} which is known to have an extremely large nonlinear index of refraction. 

We first define a bulk-equivalent susceptibility as $\chi^{(3)}_{\rm eq}=\chi^{(3){\rm 2D}}/t_{\rm 2D}$ where $t_{\rm 2D}$ is the physical thickness of the monolayer. For instance then ${\rm MoS}_2$ at 577nm and 1560nm respectively exhibits an expected bulk-equivalent, absolute nonlinear Kerr susceptibility $\chi^{(3)}_{\rm eq}(-\omega;\omega;-\omega;\omega)$ of $2.11\times 10^{-13}{\rm m^2 V^{-2}}$ and $1.17\times 10^{-12}{\rm m^2 V^{-2}}$. In a similar manner, the corresponding expected bulk-equivalent, absolute third-harmonic susceptibilities $\chi^{(3)}_{\rm eq}(-3\omega;\omega;\omega;\omega)$ are $2.96\times 10^{-17}{\rm m^2 V^{-2}}$ and $2.11\times 10^{-15}{\rm m^2 V^{-2}}$ at 577nm and 1560nm respectively. 

These numbers could reasonably well explain the recently observed ultrastrong high harmonic generation in 2D TMDCs \cite{5,6,8}. There are certain classes of materials or media, which could offer significantly stronger nonlinearity, however, either their slow response times (such as polymers) or complexity of formation (such as cold atomic gases), render them of limited use at optical frequencies.

The unusually large nonlinearity of 2D materials compared to bulk 3D structures is hard to explain. However, we believe that it is a matter of geometrical confinement dimensions that sets this strength. At least it has been rigorously established that the third order nonlinear optical response of spin density wave insulators is much stronger in 2D than 3D and 1D structures \cite{6a}.

\subsubsection{Analogy with Experimental Values}

The validity of numerical results could be furthermore verified against three recent experimental data \cite{5,6,5a} on 3rd harmonic generation from ${\rm MoS}_2$ at the wavelength of 1560nm. While our developed code estimates a value of $1.6\times 10^{-25} {\rm m^3 V^{-2}}$ for $\chi_{mnpq}^{(3)}(-3\omega;\omega,\omega,\omega)$, as shown in Fig. \ref{Fig5}, the reported experimental data are $3.9\times 10^{-15}{\rm m^2 V^{-2}}\times 0.75{\rm nm}=2.93\times 10^{-24}{\rm m^3 V^{-2}}$ \cite{6}, as well as the very different values of $1.7\times 10^{-28} {\rm m^3 V^{-2}}$ \cite{5}, as well as $1.2\times 10^{-19}{\rm m^2 V^{-2}}\times 0.65{\rm nm}=7.8\times 10^{-29} {\rm m^3 V^{-2}}$ as the pre-treatment and $1.95\times 10^{-28} {\rm m^3 V^{-2}}$ as the post-treatment values \cite{5a}, and also $6.5\times10^{-29}{\rm m^3 V^{-2}}$ \cite{5b}. 

This shows that while experimental results \cite{5,6,5a} vary within four orders of magnitude, our numerical estimate is much closer to the first measurement \cite{6} reporting the larger value. The difference between theoretical results with experimental ones, therefore, should not be surprising considering the remarkably scattered numerical values among experimental observations. While the nature of such differences is not exactly known yet, it could be due to material growth and transfer issues, defect concentrations, substrate effects, as well as the optical method of measurement. Given the large sensitivity of TMDCs to the fabrication process and even resilience under ambient conditions, there remains a question of how to unify experiments alike. It could be speculated that for a variety of reasons, the experimental reports actually either underestimate or overestimate the nonlinear susceptibility over the true theoretical value. 

\subsubsection{Behavior of Tensor Components}

With regard to the individual tensor components recognized by the set of indices $\mu\nu\zeta\eta$, there should be $2^4=16$ elements. However, half of them are zero and the only non-zero tensor elements are yyyy, xxxx, xyxy, yxyx, xyyx, yxxy, xxyy, and yyxx. Even though, all tensor elements are not quite independent and the following identities hold for all four TMDCs due to the crystal symmetries
\begin{eqnarray}
\label{eqa}
\chi^{(3)}_{yyyy}&&=\chi^{(3)}_{xxxx}=\chi^{(3)}_{yyxx}+\chi^{(3)}_{yxxy}+\chi^{(3)}_{yxyx},\\ \nonumber
\chi^{(3)}_{yyxx}&&=\chi^{(3)}_{xxyy},\\ \nonumber
\chi^{(3)}_{yxxy}&&=\chi^{(3)}_{xyyx},\\ \nonumber
\chi^{(3)}_{yxyx}&&=\chi^{(3)}_{xyxy}.
\end{eqnarray}
\noindent
This limits the maximum number of independent tensor elements to three. 

For the case of Kerr nonlinearity where the parameter $\chi^{(3)}_{\mu\nu\zeta\eta}(-\omega;\omega,-\omega,\omega)$ is concerned, one may verify that the elements yyxx and yxyx also identically vanish. Hence, we have $\chi^{(3)}_{yyyy}=\chi^{(3)}_{xxxx}=\chi^{(3)}_{yxxy}=\chi^{(3)}_{xyyx}$ and get a fairly simple scalar form for the nonlinear Kerr susceptibility as
\begin{equation}
\label{eqb}
\chi^{(3)}_{\mu\nu\zeta\eta}(-\omega;\omega,-\omega,\omega)=\chi^{(3)}(\omega)\delta_{\mu\eta}\delta_{\nu\zeta}.
\end{equation}

Similarly, we could write for the third-harmonic generation the following
\begin{eqnarray}
\label{eqc}
\chi^{(3)}_{\mu\nu\zeta\eta}&&(-3\omega;\omega,\omega,\omega)=\chi^{(3)}_1(\omega)\delta_{\mu\nu}\delta_{\zeta\eta}\\ \nonumber
&&+\chi^{(3)}_2(\omega)\delta_{\mu\eta}\delta_{\nu\zeta}+\chi^{(3)}_3(\omega)\delta_{\mu\zeta}\delta_{\nu\eta},
\end{eqnarray}
\noindent
where $\chi^{(3)}_l(\omega)$ with $l=1,2,3$ are independent scalars.

\section{Non-classical State of Light}

In this section, we discuss production of a non-classical state of light with elliptical Wigner distribution, due to two-photon absorption. Since the method of analysis is based on the theory of squeezing, we need to initially present an overview of light squeezing schemes.

\subsection{Methods of Squeezing}

Squeezing of light is usually done through either of the following general methods \cite{9,10,Sq1,Sq2,Sq3,Sq4,Sq4a}:

\begin{itemize}
	\setlength\itemsep{0mm}
	\item Squeezed light by parametric down conversion,
	\item Squeezed light in optical fibers,
	\item Squeezed light in atomic ensembles, and
	\item Squeezed light in semiconductor lasers.
\end{itemize}

Out of the above four methods, the first three are mostly implemented in a non-monolithic experiments and normally require large optical setups. The second one is based on the $\chi^{(3)}$ effect of fused silica in optical fibers, and thus requires long propagation paths over fibers to achieve squeezing, but squeezing as large as 9dB has been achieved using this scheme. The third one requires sophisticated techniques of atom vapor trapping and condensation at ultralow temperatures, and the largest observed squeezing of 13dB has been so far achieved this way. The fourth method \cite{Sq10} is relatively easy to achieve and based on monolithic integrated photonics, but is limited for various practical reasons to only 3-4dB of reduction in shot noise and thus squeezing. 

Recent advances in optomechanics, has brought the possibility of optomechanical squeezing of light into perspective as well \cite{11,12,12a,12b,12c,12d}. Optomechanical squeezing of light in homodyne detection within a small amount also occurs, and can be observed using a novel quantum feedback control scheme which has been recently reported \cite{13}. Monolithic approaches to squeezing by optical parametric oscillators \cite{14} have been shown to be feasible as well.

As an alternative route, the possibility of using ${\rm Si_3N_4}$ micro-ring resonators for squeezing has been demonstrated in a multi-mode optical parametric oscillator \cite{15} with 1.7dB squeezing. This has also apparently been verified experimentally at room temperature by pumping a continuous laser \cite{16}, and 0.5dB squeezing below shot noise was observed in a self-homodyne setup. So, it could be safely claimed that using silica microdisks covered with 2D TMDCs, certainly measurable squeezing could be obtained. It has been furthermore recently shown that high-quality silica micro-ring resonators could provide a versatile platform for study of emission properties of 2D materials \cite{Ex1}.

\subsection{$\chi^{(3)}$ Squeezing on Micro-Resonators}

It is known that the 3rd order nonlinearity could be employed to generate squeeze light. Depending on the optical setup, that whether the squeezed state is produced through unitary transformation in vacuum, or an interferometric setup, it leads to either of the  Hamiltonians \cite{1,2} as
\begin{eqnarray}
\label{eq7a}
\mathbb{H}&=&\hbar\omega\left(\hat{a}^\dagger\hat{a}+\frac{1}{2}\right)+\hbar\omega\left(\xi\hat{a}^{\dagger 2}-\xi^*\hat{a}^2\right),\\ 
\label{eq7b}
\mathbb{H}&=&\hbar\omega\left(\hat{a}^\dagger\hat{a}+\frac{1}{2}\right)+\hbar\omega\xi\hat{a}^{\dagger 2}\hat{a}^2.
\end{eqnarray}
The first squeezed state (\ref{eq7a}) being referred to as the quadrature squeezing, has an elliptical Wigner distribution, while the second one (\ref{eq7b}) being referred to as the photon-number squeezing results in a  kidney- or crescent-shaped squeezing and is thus quite different and non-Hermitian when $\Im[\xi]\neq 0$. Here, $\omega$ is the frequency of light and $\xi$ is a dimensionless parameter defined as \cite{2}
\begin{equation}
\label{eq8}
\xi=\frac{3\hbar\omega\mathcal{F}}{8\epsilon_0}\sum_{mnpq}\iiint\chi^{(3)}_{mnpq}({\bf r})u_m({\bf r})u_n({\bf r})u_p^*({\bf r})u_q^*({\bf r})d^3r,
\end{equation}
\noindent
where $\textbf{u}({\bf r})$ is the vector mode profile of a single photon in the cavity with components $u_m({\bf r})$, and a factor $\frac{3}{2}$ is already included for standing wave, as it is going to be the case under study. Furthermore, since the optical wave is strongly confined in the micro-disk and undergoes a much larger effective propagation path, an extra dimensionless finesse factor $\mathcal{F}$ is also included. The finesse $\mathcal{F}$ is a direct measure of how many times the light pulse circulates the ring \cite{Fin1,Fin2}. This is while the single photon mode $\textbf{u}({\bf r})$ having the physical dimension ${\rm m}^{-3/2}$ must be normalized as 
\begin{eqnarray}
\label{eq9}
1&=&\sum_{mn}\iiint\epsilon_{r,mn}({\bf r})u_m({\bf r})u_n^*({\bf r})d^3r\\ \nonumber
&=&\sum_{mn}\iiint\left[\delta_{mn}+\chi^{(1)}_{mn}({\bf r})\right]u_m({\bf r})u_n^*({\bf r})d^3r.
\end{eqnarray}
\noindent
Here, $\chi^{(1)}({\bf r})=\epsilon_r({\bf r})-1$ is the relative susceptibility of the micro-disk resonator, including the cladding and substrate. Hence, (\ref{eq9}) is effectively taken on the mode volume of the cavity. Obviously, (\ref{eq8}) could be written for an unnormalized mode as
\begin{eqnarray}
\label{eq10}
\xi&&=\frac{3\hbar\omega\mathcal{F}}{8\epsilon_0}\times\\ \nonumber
&&\frac{\sum_{mnpq}\iiint\chi^{(3)}_{mnpq}({\bf r})u_m({\bf r})u_n({\bf r})u_p^*({\bf r})u_q^*({\bf r})d^3r}{\left(\sum_{mn}\iiint\left[\delta_{mn}+\chi^{(1)}_{mn}({\bf r})\right]u_m({\bf r})u_n^*({\bf r})d^3r\right)^2}.
\end{eqnarray}
In presence of a 2D material, and using a conducting interface approximation \cite{4a}, these two latter expressions should be changed slightly as follows
\begin{eqnarray}
\label{eq11a}
\xi&=&\frac{3\hbar\omega\mathcal{F}}{8\epsilon_0}\times\sum_{mnpq}\\ \nonumber
&&\Bigg\{\iiint_{\rm Disk}\chi^{(3)}_{mnpq}({\bf r})u_m({\bf r})u_n({\bf r})u_p^*({\bf r})u_q^*({\bf r})d^3r,\\ \nonumber
&+&\iint_{\rm 2D}\chi^{(3){\rm 2D}}_{mnpq}({\bf r})u_m({\bf r})u_n({\bf r})u_p^*({\bf r})u_q^*({\bf r})d^2r\Big|_{z=0}\Bigg\},\\ 
\label{eq11b}
1&=&\sum_{mn}\iiint_{\rm Disk}\left[\delta_{mn}+\chi^{(1)}_{mn}({\bf r})\right]u_m({\bf r})u_n^*({\bf r})d^3r,\\ \nonumber
&+&\sum_{mn}\iint_{\rm 2D}\chi^{(1){\rm 2D}}_{mn}({\bf r})u_m({\bf r})u_n^*({\bf r})d^2r\Big|_{z=0}.
\end{eqnarray}
\noindent
where it is supposed that the 2D TMDC is placed on the $z=0$ plane. Following (\ref{eqb}) for Kerr nonlinearity and two-photon absorption in unstrained TMDCs, both $\chi^{(3)}$ and $\chi^{(1)}$ are scalar quantities. This greatly simplifies (\ref{eq11a},\ref{eq11b}) as
\begin{eqnarray}
\label{eq12a}
\xi&=&\frac{3\hbar\omega\mathcal{F}}{8\epsilon_0}\Bigg\{\iiint_{\rm Disk}\chi^{(3)}({\bf r})|\textbf{u}({\bf r})|^4d^3r\\ \nonumber
&+&\iint_{\rm 2D}\chi^{(3){\rm 2D}}({\bf r})|\textbf{u}({\bf r})|^4d^2r\Big|_{z=0}\Bigg\},\\ 
\label{eq12b}
1&=&\iiint_{\rm Disk}\left[1+\chi^{(1)}({\bf r})\right]|\textbf{u}({\bf r})|^2d^3r\\ \nonumber
&+&\iint_{\rm 2D}\chi^{(1){\rm 2D}}({\bf r})|\textbf{u}({\bf r})|^2d^2r\Big|_{z=0}.
\end{eqnarray}

\subsubsection{Effective Nonlinearity \& Mode Volume}

Alternatively, (\ref{eq12a}) can be written as 
\begin{equation}
\label{eq13a}
\xi=\frac{3\hbar\omega\mathcal{F}}{8\epsilon_0V}\chi_{\rm eff}^{(3)},
\end{equation}
\noindent 
where $V$ is mode volume \cite{Sq13d} and $\chi_{\rm eff}^{(3)}$ is the effective nonlinear index defined as
\begin{eqnarray}
\label{eq14a}
V&=&\left[\iiint_{\rm Disk}|\textbf{u}({\bf r})|^4d^3r\right]^{-1}, \\ \nonumber
\chi_{\rm eff}^{(3)}&=&V\Bigg\{\iiint_{\rm Disk}\chi^{(3)}({\bf r})|\textbf{u}({\bf r})|^4d^3r\\ \nonumber
&+&\iint_{\rm 2D}\chi^{(3){\rm 2D}}({\bf r})|\textbf{u}({\bf r})|^4d^2r\Big|_{z=0}\Bigg\}.
\end{eqnarray}

It should be mentioned here that $\xi$ and therefore $\chi_{\rm eff}^{(3)}$ by definition cannot be independent of the micro-disk radius $r$ as well as wavelength $\lambda$.

Another way to look into this is to view the circulating optical power in disk resonator as an optical field going through a long straight path \cite{Sq5,Sq6}, thus experiencing an overall phase retardation. This is the basis of light squeezing in optical fibers \cite{8a,Fib1,Fib2,Fib3}, and this point of view gives a more clear and straightforward measure for the squeezing parameter $s$ as
\begin{eqnarray}
s&=&\frac{6\pi r\epsilon_0\mathcal{F}}{n\lambda(N\hbar\omega)}\Bigg|\iiint_{\rm Disk}\chi^{(3)}({\bf r})|\textbf{E}({\bf r})|^4d^3r\\ \nonumber
&+&\iint_{\rm 2D}\chi^{(3){\rm 2D}}({\bf r})|\textbf{E}({\bf r})|^4d^2r\Big|_{z=0}\Bigg|,
\end{eqnarray}
\noindent
where $r$ and $n$ are disk's radius and index of refraction, respectively, $\lambda$ is the optical wavelength, and $\textbf{E}({\bf r})$ is the electric field inside the disk. Moreover, $N$ is the number of photons inside the disk, and thus $N\hbar\omega$ is the total optical energy confined in the cavity. 

Now, plugging in (\ref{eq12a}) and further simplification gives the final expression as
\begin{eqnarray}
s&=&\frac{8\epsilon_0^2 rV}{n\hbar c(N\hbar\omega)}\Bigg|\xi\iiint_{\rm Disk}\bar{\chi}^{(3)}({\bf r})|\textbf{E}({\bf r})|^4d^3r\\ \nonumber
&+&\iint_{\rm 2D}\bar{\chi}^{(3){\rm 2D}}({\bf r})|\textbf{E}({\bf r})|^4d^2r\Big|_{z=0}\Bigg|.
\end{eqnarray}
\noindent
Here, $\bar{\chi}^{(3){\rm 2D}}=\bar{\chi}^{(3){\rm 2D}}/\chi^{(3)}_{\rm eff}$ and $\bar{\chi}^{(3)}=\bar{\chi}^{(3)}/\chi^{(3)}_{\rm eff}$ are dimensionless quantities. It should be noted that in the above relations, $N\hbar\omega$ is in denominator, but since $|\textbf{E}|^2\propto N\hbar\omega$, the overall squeeze factor is proportional to the optical energy inside the cavity. More specifically, we have $\textbf{E}=\sqrt{N\hbar\omega/2\epsilon_0}\textbf{u}$ \cite{10}, which gives the simple form
\begin{equation}
s=\frac{4\pi r}{\lambda}|\xi| N.
\end{equation}

\subsubsection{Noise Squeezing/Desqueezing}

As it was discussed in the above, $\chi^{(3)}$ is purely imaginary in the wavelength range of interest between 760nm to 790nm, and thus it is the two-photon absorption which wins over the Kerr nonlinear index. Despite this fact, this is unimportant for our particular application of producing a non-classical state of light with non-circular Wigner distribution. 

Some studies have only considered the absolute value of $|\chi^{(3)}|$ to  the squeeze parameter $s=|\zeta|$ \cite{8a}, where $\zeta$ is the complex squeeze parameter given by the squeeze operator as \cite{8a,8b,8c,8d,10}
\begin{equation}
\label{sq1}
\hat{S}(\zeta)=\exp\left(\frac{1}{2}\zeta^*\hat{a}^2-\frac{1}{2}\zeta\hat{a}^{\dagger 2}\right).
\end{equation}
\noindent
Ignoring the real part of $\zeta$ and noting that its imaginary part is positive, we may rewrite the above as
\begin{equation}
\label{sq2}
\hat{S}(s)=\exp\left[\frac{-is}{2}\left(\hat{a}^2+\hat{a}^{\dagger 2}\right)\right].
\end{equation}
\noindent
This is actually directly corresponding to the interaction Hamiltonian (\ref{eq7a}) in the above, noting that $s\propto|\xi |$. As it will be discussed later below, the fact that $\chi^{(3)}$ is imaginary does not disallow production of non-classical states.

It should be mentioned that actual ratio of noise squeezing for the case of (\ref{eq7a}), here being denoted respectively by $\varrho$ and $\varphi$ for the two orthogonal quadratures is not the same as $s$, since $\zeta$ is not real valued. As we can actually show in Appendix \ref{AppC}, when $\Re[\zeta]=0$ the correct relationship is
\begin{eqnarray}
\varrho&=&\cosh(s)\sqrt{1+e^{-2s}\sinh^2(s)},\\ \nonumber
\varphi&=&\sqrt{\frac{\sinh^2(2s)}{4e^{2s}}+{\rm sech}^2(s)\left[1+\frac{\sinh^2(2s)}{4e^{2s}}\right]^2}.
\end{eqnarray}
\noindent
These expressions behave as $\varrho\approx2\varphi\approx (\sqrt{5}/4)e^{s}$ in the limit of large $s$, implying asymmetric noise increase at high intensities, while approaches unity for small $s$ implying no non-classical squeezing effect at low intensities. 

As it has been demonstrated in the Appendix \ref{AppC}, the measure of non-classicality of the resulting Wigner distribution is limited to 6.02dB at high intensities or high $\chi^{(3)}_{\rm eff}$.

\subsection{Experimental Issues}

One therefore may produce a non-classical light using a micro-disk resonator covered with a 2D TMDC, such as ${\rm WSe}_2$, or ${\rm MoS}_2$. We would like to investigate this possibility by taking advantage of the very large $\chi^{(3)}$ coefficient of 2D TMDCs. Combined with the very strong confinement of light in low loss silica disk resonators, we would expect a significant anharmonicity $\xi$ as defined in (\ref{eq11a}). This has yet to be calculated. If $\xi$ is hopefully found to be reasonably large, and leading to an observable value of desqueezing asymmetry in light, we would move ahead with experiments to characterize the properties of output emission and shot noise. The enhanced effective nonlinear susceptibility $\chi^{(3)}_{\rm eff}$ should in principle allow production of non-classical elongated state at much lower intensities.

\subsubsection{Ultrashort Solitons}

A possible advantage of this scheme in case of successful design and experiments, would be relatively ease of fabrication and operation, as well as compatibility with monolithic integrated photonics. Usage of ultrashort solitons \cite{Sq9,Sq9a} together with strong confinement in micro-disks increases the overall nonlinear interactions and thereby squeezing as well. This has been already demonstrated in squeezing via optical fibers \cite{9}. In that case, no extra pumping is needed, and the pulse undergoes squeezing by itself through self-pumping. Evidently, the squeezed output is pulsed. Since the quality factors of cavities is not infinite, a small dissipation at high optical power densities is unavoidable \cite{Sq12,Sq13a,Sq13b,Sq13c,Sq13d}, which can be treated anyhow as either a Lugiato-Lefever equation \cite{Sq13e} or using a perturbative expansion \cite{Sq13}.

\subsubsection{Separate Pumping}

It is in practice beneficial if the optical power required for excitation of nonlinearity is maintained by a pump held at a slightly different frequency such as $\omega_p=\omega\pm{\rm FSR}$, where FSR is the free spectral range of the micro-disk resonator. The susceptibility then should be calculated from the non-resonant near-degenerate pumped value $\chi^{(3)}(-\omega;\omega_p,-\omega_p,\omega)$. Since ${\rm FSR}<<\omega$, then $\chi^{(3)}(-\omega;\omega_p,-\omega_p,\omega)$ stays within the same order of magnitude of $\chi^{(3)}(-\omega;\omega,-\omega,\omega)$, thus leaving the presented analysis and discussions basically intact. A closely related scheme with two pumps with frequencies $\omega_p^\pm=\omega\pm\Delta\Omega$ where $\Delta\Omega$ is a non-zero integer multiple of FSR has been proposed in fiber \cite{Sq7,Sq11} and recently used in integrated \cite{Sq12} ring resonators for squeezing and random number generation, respectively. It is worthwhile to mention that the reverse of this scheme where two identical pump photons decompose into two different signal and idler photons in higher and lower frequency neighboring sidebands has been shown \cite{15} to be useful for on-chip optical squeezing, too.

\begin{figure}
	\centering
	\includegraphics[width=2.2in]{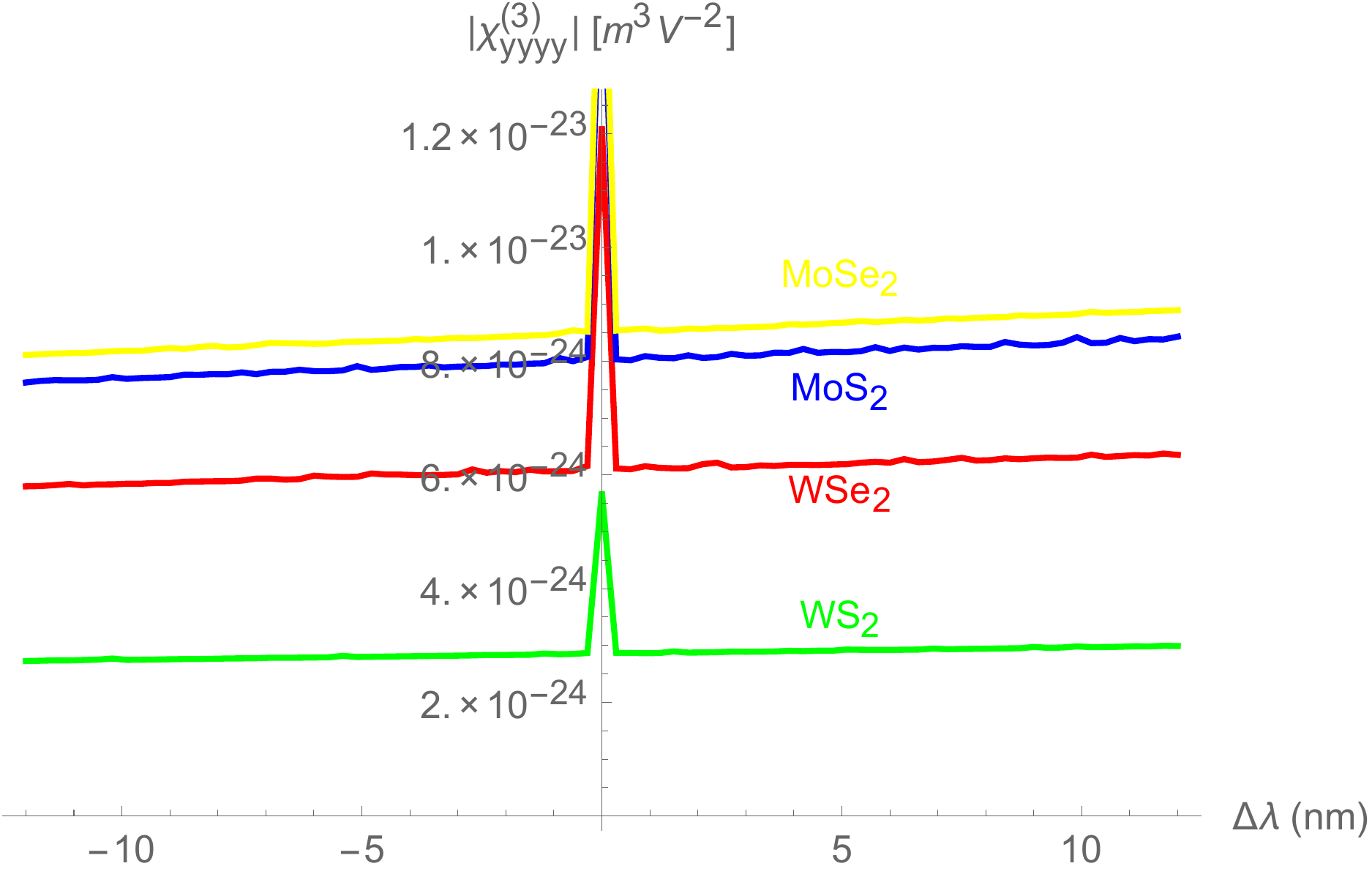}
	\caption{Computation of near-dengenerate susceptibility $\chi^{(3)}(-\omega;\omega_p,-\omega_p,\omega)$ versus pump wavelength difference $\Delta\lambda=\lambda_p-\lambda$ at fixed $\lambda=775{\rm nm}$ for various TMDCs.\label{Fig6}}
\end{figure}

A numerical evaluation of near-degenerate $\chi^{(3)}(-\omega;\omega_p,-\omega_p,\omega)$, as shown in Fig. \ref{Fig6} for an FSR of 3nm at the wavelength of 775nm, for instance, in case of ${\rm WSe}_2$ reveals that $\chi^{(3)}$ is actually reduced from the degenerate value of $|\chi^{(3)}|=1.2\times 10^{-23}{\rm m^3/V^2}$ roughly by 49.3\%, to the new non-degenerate pumped value of $|\chi^{(3)}|=6.05\times 10^{-24}{\rm m^3/V^2}$ at a pump wavelength of 778nm, which is again purely imaginary. This drop could be easily accounted for by a proportional increase in pump power. Similarly, if pumping is done at 772nm, then $|\chi^{(3)}|=5.95\times 10^{-24}{\rm m^3/V^2}$, which is a 50.2\% reduction again. Therefore, when $\Delta\lambda=\lambda_p-\lambda$ with $\lambda_p$ being the pump wavelength is sufficiently small, then the approximation 
\begin{equation}
\chi^{(3)}(-\omega;\omega_p,-\omega_p,\omega)\approx\frac{1}{2}\chi^{(3)}(-\omega;\omega,-\omega,\omega),
\end{equation}
\noindent
holds. This type of behavior is more or less the same for all other sorts of TMDCs, as shown in Fig. \ref{Fig6}.

This scheme, as opposed to the ultrashort solitons, enables continuous pumping and signal feeding, and therefore a continuous squeezed output may be expected as well.

\section{Excitonic Effects}

Without doubt, the prominent role of excitons in light emission from TMDCs could be considered as a subject of deep study. This is due to the fact that all of the four basic types of TMDCs discussed in this paper support both of the dark and bright excitons, which is further complicated by the trignoal warping property of these materials and presence of charged excitons (also known as trions) and biexcitons \cite{Ex1,ExRev1,Mueller2}. 

In general, exciton binding energies in 2D materials cannot be explained by a simple hydrogenic model, because of the very different radial distribution of the wavefunction, which could easily extend to a few nm in radius. It only can be investigated by DFT GW-BSE calculations, which exhibits numerous lines between 1s and 2p states. While the exciton binding energies in 2D materials can be large, these are directly dependent on the substrate screening effects. When there are insulating encapsulation or separation, for instance, using bilayer graphene (or BN in other works), the exciton binding energy may drastically reduce by a factor of 3 to 6. 

At room temperature, exciton emission peaks are significantly broadened, because of large electron-phonon coupling rates in typical TMDCs. Biexcitons in 2D materials are unstable at room temperature and dissociate, since their binding energies are on the order of only a few meV. The exciton-exciton annihilation (EEA) which is a four-particle process can occur under high exciton population (demanding high illumination intensities) in principle even at the room temperature, but it can be expected that at such high rates, non-radiative recombinations through defect states and impurities would be dominant over the entire EEA process. Dark excitons in most 2D materials can actually emit light. Since they are optically forbidden because of vanishing dipole. However, the true dipole should be complex valued instead of a real one, so that they can emit non-linearly polarized light (not necessarily exactly circular). The fact that the energy different between dark and bright excitons is on the same order of exciton binding energy complicates the correct interpretation of light emission.

While the existence of biexcitons at room temperature is unlikely because of the small binding energy, there are strong reasons to believe that emission from charged excitons could be easily mistaken with defect emissions because of interface effects \cite{Urb1,Exciton3,Exciton4}. Quite possibly, the existence of higher-wavelength emission peaks in photo-luminescence spectra of TMDCs could be due completely due to interface defects, and in that case trions/biexciton emissions can have no influence at room-temperature as they normally appear only at cryogenic conditions \cite{Aha}. If ture, then the observed trion emission peaks could be actually a defect-related emission, and for this reason is entirely absent in boron-nitride (BN) encapsulated monolayer TMDCs \cite{Urb1,Aha}. Similarly, there exists experimental evidence for brightened dark excitons/trions in ${\rm WSe}_2$, where measurements are all done at low temperatures \cite{Ex7}. 

Another recent study \cite{Ex10} is citing the fact that the two emission peaks for $\text{MoS}_2$ on thermally grown $\text{SiO}_2$, correspond to A-exciton and interface Defect. The defect state disappears at room temperature, which is somehow connected to observations reported in \cite{Urb1}. Furthermore, they propose an elegant way to identify a defect state.

The emission of a defect is almost always unpolarized regardless of the pump incidence angle, intensity, and polarization, and that is why it cannot form polaritons in cavity quantum electrodynamics (CQED) experiments. A careful polarization measurement on the emission spectra could unanimously reveal whether the lower energy peak is a defect or excitonic emission.

In summary, the origin and nature of peaks in the emission spectra of TMDCs has been a matter of unresolved debate \cite{Ex2,Ex3,Ex4,Ex5,Ex6,Ex7,Ex8,Ex9,Ex10,Aha,Urb1,Exciton1,Exciton2,Exciton3,Exciton4,Exciton5,Exciton6,Exciton7,Exciton8} and still remains with no conclusive agreement so far. 

\section{Conclusions}

We presented a detailed theoretical analysis of linear and third-order nonlinear optical response of two-dimensional (2D) monolayer transition metal dichalgonides (TMDCs). Based on a rigorous six-band tight-binding model, we calculated first order and third order susceptibility tensors, and observed reasonable consistency with experimental data. We predict an elongated non-classical light using high quality silica micro-disk resonators covered with the monolayer TMDCs.

\begin{acknowledgments}
	
This work has been supported in part by Laboratory of Photonics and Quantum Measurements (LPQM) at EPFL and European Graphene Flagship under grant 696656, as well as Research Deputy of Sharif University of Technology.

The author would like to thank Dr. Habib Rostami at Istituto Italiano di Tecnologia (IIT) as well as Dr. Rafael Rold\'{a}n and Dr. Pablo San-Jose at Instituto de Ciencia de Materiales de Madrid (CSIC) for discussions and assistance on the Tight-Binding method. Enlightening discussions with Prof. Steven Louie from University of California at Berkeley, Dr. Bernhard Urbaszek at Laboratoire de Physique et Chimie des Nano-objets, Dr. Igor Aharonovich at University of Technology Sydney, as well as Dr. Anshuman Kumar and Cl\'{e}ment Javerzac-Galy at \'{E}cole Polytechnique F\'{e}d\'{e}rale de Lausanne (EPFL) on excitonic emission effects is highly appreciated. 

The author is highly indebted to Hiwa Mahmoudi at Institute of Electrodynamics, Microwave and Circuit Engineering in Technische Universit\"{a}t Wien, and in particular, the Laboratory for Quantum Foundations and Quantum Information on the Nano- and Microscale in Vienna Center for Quantum Science and Technology (VCQ) at Universit\"{a}t Wien for their warm and receptive hospitality during the preparation of this article. 

This paper is dedicated to the celebrated artist, Anastasia Huppmann.

\end{acknowledgments}

\appendix
\section*{Appendices}
\setcounter{section}{0}

\section{Six-Band Tight-Binding}\label{AppA}

The $6\times 6$ Hamiltonian can be decomposed as \cite{3c6a,3c6b,3c6c,3c6d}
\begin{eqnarray}
\mathbb{H}({\bf k};\sigma)&=&\begin{bmatrix}
\mathbb{H}_{\rm MM}({\bf k};\sigma) & \mathbb{H}_{\rm MX}({\bf k})\\
\mathbb{H}_{\rm MX}^\dagger({\bf k}) & \mathbb{H}_{\rm XX}({\bf k};\sigma)
\end{bmatrix},\\ \nonumber
\mathbb{H}_{\rm MM}({\bf k};\sigma)&=& \mathbb{E}_{\rm M}(\sigma)+2\sum^{3}_{j=1}\mathbb{T}_j^{\rm MM}\cos({\bf k}\cdot{\bf a}_j),\\ \nonumber
\mathbb{H}_{\rm XX}({\bf k};\sigma)&=& \mathbb{E}_{\rm X}(\sigma)+2\sum^{3}_{j=1}\mathbb{T}_j^{\rm XX}\cos({\bf k}\cdot{\bf a}_j),\\ \nonumber
\mathbb{H}_{\rm MX}({\bf k})&=& \sum^{3}_{j=1}\mathbb{T}_j^{\rm MX}\exp(-i{\bf k}\cdot\boldsymbol{\delta}_j). 
\end{eqnarray}
\noindent
The vectors ${\bf a}_j$ form the lattice basis vectors and extend from every metal to the nearest neighbors. They are all equal in length, given as the lattice constant $a=|{\bf a}_j|$. The vectors $\boldsymbol{\delta}_j$ have equal length, given by $|\boldsymbol{\delta}_j|=\delta=a/\sqrt{3}$, but make right angles with the basis vectors. A desirable choice for these vectors are ${\bf a}_1=\frac{1}{2}a(-1,\sqrt{3})$, ${\bf a}_3=\frac{1}{2}a(-1,-\sqrt{3})$, and ${\bf a}_2=a(1,0)$. Then we have $\boldsymbol{\delta}_1=\frac{1}{2}\delta(\sqrt{3},-1)$, $\boldsymbol{\delta}_3=\frac{1}{2}\delta(-\sqrt{3},-1)$, and $\boldsymbol{\delta}_2=\delta(0,1)$. With these conditions, the coordinates of high-symmetry reciprocal lattice points are ${\rm K}=\frac{4\pi}{3a}(1,0)$, ${\rm M}=\frac{\pi}{3\sqrt{3}}(\sqrt{3},1)$, and $\Gamma=(0,0)$. In Table \ref{Tab1}, the lengths of these vectors for various TMDCs are introduced.

\begin{table}[h]
	\centering
	\caption{Lattice constants $a=\sqrt{3}\delta$ of TMDCs \cite{3c6d}.}
	\label{Tab1}
	\begin{tabular}{llll}
		\hline\hline
		${\rm MoS}_2$ & ${\rm MoSe}_2$  & ${\rm WS}_2$  & ${\rm WSe}_2$ \\ \hline
		3.160\AA & 3.288\AA  & 3.153\AA  & 3.260\AA  \\ \hline\hline
	\end{tabular}
\end{table}

\begin{table}[h]
	\centering
	\caption{Slater-Koster parameters for TMDCs. All parameters are in units of electron-volts (adapted from \cite{3c6d} with minor corrections).}
	\label{Tab2}
	\begin{tabular}{llllll}
		\hline\hline
		&  & ${\rm MoS}_2$ & ${\rm MoSe}_2$ & ${\rm WS}_2$ & ${\rm WSe}_2$ \\ \hline
		\multirow{2}{*}{SOC} & $\lambda_{\rm M}$  & 0.086 & 0.089 & 0.271 & 0.251 \\
		&  $\lambda_{\rm X}$ & 0.052 & 0.256 & 0.057 & 0.439 \\ \hline
		\multirow{5}{*}{Crystal Fields} & $\Delta_0$ & $-1.094$ & $-1.144$ & $-1.155$ & $-0.935$ \\
		& $\Delta_1$ & $-0.050$ & $-0.250$ & $-0.650$ & $-1.250$ \\
		& $\Delta_2$ & $-1.511$ & $-1.488$ & $-2.279$ & $-2.321$ \\
		& $\Delta_p$ & $-3.559$ & $-4.931$ & $-3.864$ & $-5.629$ \\
		& $\Delta_z$ & $-6.886$ & $-7.503$ & $-7.327$ & $-6.759$ \\ \hline
		\multirow{2}{*}{${\rm M-X}$} & $V_{pd\sigma}$  & 3.689  & 3.728 & 4.911 & 5.083 \\
		& $V_{pd\pi}$ & $-1.241$ & $-1.222$ & $-1.220$  & $-1.081$ \\ \hline
		\multirow{3}{*}{${\rm M-M}$} & $V_{dd\sigma}$ & $-0.895$  & $-0.823$ & $-1.328$ & $-1.129$ \\
		& $V_{dd\pi}$ & 0.252 & 0.215 & 0.121 & 0.094 \\
		& $V_{pd\delta}$ & 0.228 & 0.192 & 0.442 & 0.317 \\ \hline
		\multirow{2}{*}{${\rm X-X}$} & $V_{pp\sigma}$ & 1.225 & 1.256 & 1.178 & 1.530 \\
		& $V_{pp\pi}$ & $-0.467$ & $-0.205$ & $-0.273$ & $-0.123$ \\ \hline\hline
	\end{tabular}
\end{table}
The $3\times 3$ submatrices in the above are given as
\begin{eqnarray}
\mathbb{E}_{\rm M}(\sigma)&=&\begin{bmatrix}
\Delta_0 & 0 & 0 \\
0 & \Delta_2 & -i\lambda_{\rm M}\sigma \\
0 & i\lambda_{\rm M}\sigma & \Delta_2
\end{bmatrix}, \\ \nonumber
\mathbb{E}_{\rm X}(\sigma)&=&\begin{bmatrix}
\Delta_p+V_{pp\pi} & -\frac{i}{2}\lambda_{\rm X}\sigma & 0 \\
\frac{i}{2}\lambda_{\rm X}\sigma & \Delta_p+V_{pp\pi} & 0 \\
0 & 0 & \Delta_z-V_{pp\sigma}
\end{bmatrix}.
\end{eqnarray}

The rest of the matrices are introduced as follows. Starting with $\mathbb{T}_j^{\rm MM}$ we have

\begin{widetext}
	\begin{equation}
	\mathbb{T}_1^{\rm MM}=\frac{1}{4}\begin{bmatrix}
	3V_{dd\delta}+V_{dd\sigma} & \frac{\sqrt{3}}{2}(-V_{dd\delta}+V_{dd\sigma}) & -\frac{3}{2}(V_{dd\delta}-V_{dd\sigma})\\
	\frac{\sqrt{3}}{2}(-V_{dd\delta}+V_{dd\sigma}) & \frac{1}{4}(V_{dd\delta}+12V_{dd\pi}+3V_{dd\sigma}) & \frac{\sqrt{3}}{4}(V_{dd\delta}-4V_{dd\pi}+3V_{dd\sigma}) \\
	-\frac{3}{2}(V_{dd\delta}-V_{dd\sigma}) & \frac{\sqrt{3}}{4}(V_{dd\delta}-4V_{dd\pi}+3V_{dd\sigma}) & \frac{1}{4}(3V_{dd\delta}+4V_{dd\pi}+9V_{dd\sigma}) 
	\end{bmatrix}, 
	\end{equation}
	\begin{equation}\nonumber
	\mathbb{T}_2^{\rm MM}=\frac{1}{4}\begin{bmatrix}
	3V_{dd\delta}+V_{dd\sigma}& \sqrt{3}(V_{dd\delta}-V_{dd\sigma}) & 0 \\
	\sqrt{3}(V_{dd\delta}-V_{dd\sigma}) & 3V_{dd\delta}+V_{dd\sigma} & 0\\
	0 & 0 &  4V_{dd\pi}
	\end{bmatrix}, 
	\end{equation}
	\begin{equation} \nonumber
	\mathbb{T}_3^{\rm MM}=\frac{1}{4}\begin{bmatrix}
	3V_{dd\delta}+V_{dd\sigma} & \frac{\sqrt{3}}{2}(-V_{dd\delta}+V_{dd\sigma}) & \frac{3}{2}(V_{dd\delta}-V_{dd\sigma}) \\
	\frac{\sqrt{3}}{2}(-V_{dd\delta}+V_{dd\sigma}) & \frac{1}{4}(V_{dd\delta}+12V_{dd\pi}+3V_{dd\sigma}) & -\frac{\sqrt{3}}{4}(V_{dd\delta}-4V_{dd\pi}+3V_{dd\sigma}) \\
	\frac{3}{2}(V_{dd\delta}-V_{dd\sigma}) & -\frac{\sqrt{3}}{4}(V_{dd\delta}-4V_{dd\pi}+3V_{dd\sigma}) & \frac{1}{4}(3V_{dd\delta}+4V_{dd\pi}+9V_{dd\sigma})
	\end{bmatrix}.
	\end{equation}
	For $\mathbb{T}_j^{\rm XX}$ we have
	\begin{equation}
	\mathbb{T}_1^{\rm XX}=\frac{1}{4}\begin{bmatrix}
	3V_{pp\pi}+V_{pp\sigma}& \sqrt{3}(V_{pp\pi}-V_{pp\sigma}) & 0 \\
	\sqrt{3}(V_{pp\pi}-V_{pp\sigma})& V_{pp\pi}+3V_{pp\sigma} & 0 \\
	0 & 0 & 4V_{pp\pi}
	\end{bmatrix}, 
	\end{equation}
	\begin{equation} \nonumber
	\mathbb{T}_2^{\rm XX}=\begin{bmatrix}
	V_{pp\sigma} & 0 & 0 \\
	0 & V_{pp\pi} & 0 \\
	0 & 0 & V_{pp\pi}
	\end{bmatrix}, 
	\end{equation}
	\begin{equation} \nonumber
	\mathbb{T}_3^{\rm XX}=\frac{1}{4}\begin{bmatrix}
	3V_{pp\pi}+V_{pp\sigma}& -\sqrt{3}(V_{pp\pi}-V_{pp\sigma}) & 0 \\
	-\sqrt{3}(V_{pp\pi}-V_{pp\sigma})& V_{pp\pi}+3V_{pp\sigma} & 0 \\
	0 & 0 & 4V_{pp\pi}
	\end{bmatrix}.
	\end{equation}
	The submatrices $\mathbb{T}_j^{\rm MX}$ given in \cite{3c6d}, with minor corrections take the form
	\begin{equation}
	\mathbb{T}_1^{\rm MX}=\frac{\sqrt{2}}{7\sqrt{7}}\begin{bmatrix}
	-9V_{pd\pi}+\sqrt{3}V_{pd\sigma} & 3\sqrt{3}V_{dp\pi}-V_{pd\sigma} & -12V_{pd\pi}-\sqrt{3}V_{pd\sigma}\\
	5\sqrt{3}V_{pd\pi}+3V_{pd\sigma} & 9V_{pd\pi}-\sqrt{3} V_{pd\sigma} & 2\sqrt{3}V_{pd\pi}-3V_{pd\sigma}\\
	-V_{pd\pi}-3\sqrt{3}V_{pd\sigma} & 5\sqrt{3}V_{pd\pi} +3 V_{pd\sigma} & -6V_{pd\pi} +3\sqrt{3}V_{pd\sigma}
	\end{bmatrix},
	\end{equation}
	\begin{equation}\nonumber
	\mathbb{T}_2^{\rm MX}=\frac{\sqrt{2}}{7\sqrt{7}}\begin{bmatrix}
	0 & -6\sqrt{3}V_{pd\pi}+2 V_{pd\sigma} & -12V_{pd\pi}-\sqrt{3} V_{pd\sigma}\\
	0 & -6V_{pd\pi} -4\sqrt{3}V_{pd\sigma} & -4\sqrt{3}V_{pd\pi}+6 V_{pd\sigma}\\
	14V_{pd\pi} & 0 & 0
	\end{bmatrix}, 
	\end{equation}
	\begin{equation} \nonumber
	\mathbb{T}_3^{\rm MX}=\frac{\sqrt{2}}{7\sqrt{7}}\begin{bmatrix}
	9V_{pd\pi}-\sqrt{3} V_{pd\sigma} & 3\sqrt{3}V_{pd\pi}- V_{pd\sigma} & -12V_{pd\pi}-\sqrt{3} V_{pd\sigma} \\
	-5\sqrt{3}V_{pd\pi} -3V_{pd\sigma} & 9V_{pd\pi}-\sqrt{3} V_{pd\sigma} & 2\sqrt{3}V_{pd\pi}-3 V_{pd\sigma}\\
	-V_{pd\pi}-3\sqrt{3} V_{pd\sigma} & -5\sqrt{3}V_{pd\pi}-3 V_{pd\sigma} & 6V_{pd\pi} -3\sqrt{3}V_{pd\sigma}
	\end{bmatrix}. \nonumber
	\end{equation}
	
	Table \ref{Tab2} presents the Slater-Koster parameters needed for this analysis. Data are compiled from literature \cite{3c6d} with minor corrections, and numerical results from the model presented in this Appendix are basically identical to the ones reported therein. Comparison to experimental values are already done in many of works, however, where the interested reader is referred to.
	
	\section{Expressions for $\chi^{(3)}$}\label{AppB}
	
	Following the standard perturbation method to calculate the nonlinear susceptibility tensor \cite{6,7,7a,7b,7c,7d,7e,7f,7g,7h,7i,7j,7k}, the $\chi^{(3)}$ is composed of two paramagnetic $\Pi$ and diamagnetic $\Delta$ parts, as 
	\begin{eqnarray}
	\label{A1}
	\chi_{\mu\nu\zeta\eta}^{(3)}&&(-\omega_{\Sigma } ;\omega_r ,\omega_s , \omega_t )=\frac{6}{A_{\rm UC}\epsilon_0(\omega_{\Sigma } \omega_r \omega_s \omega_t )}\times\\ \nonumber &&\sum_{\sigma} \frac{1}{A_{\rm BZ}} \iint_{\rm IRBZ}\left[ \Pi_{\mu\nu\zeta\eta}^{\sigma}({\bf k}; \omega_r ,\omega_s ,\omega_t )+\Delta_{\mu\nu\zeta\eta}^{\sigma}({\bf k}; \omega_r ,\omega_s ,\omega_t )\right]d^2k,
	\end{eqnarray}
	\noindent
	where $\omega_\Sigma=\omega_r+\omega_s+\omega_t$. For the case of third-harmonic generation, we have $\omega=\omega_r=\omega_s=\omega_t$ and $\omega_\Sigma=3\omega$. For the Kerr nonlinearity we have $\omega=\omega_r=-\omega_s=\omega_t=\omega_\Sigma$. 
	
	For the paramagnetic contribution, we have
	\begin{eqnarray}
	\Pi_{\mu\nu\zeta\eta}^{\sigma}&&({\bf k}; \omega_r ,\omega_s ,\omega_t )=\\ \nonumber&&\mathcal{P}\sum_{mnpq} \frac{j_{mn}^{\mu\sigma}({\bf k})j_{np}^{\nu\sigma}({\bf k})j_{pq}^{\zeta\sigma}({\bf k})j_{qm}^{\eta\sigma}({\bf k})}{\hbar\omega_\Sigma+E_{mq}^\sigma({\bf k})}\Xi_{mnpq}^{\sigma}({\bf k}; \omega_r ,\omega_s ,\omega_t ),\\ 
	\Xi_{mnpq}^{\sigma}&&({\bf k}; \omega_r ,\omega_s ,\omega_t )=\\ \nonumber &&W^\sigma_{mnp}({\bf k};\omega_r+\omega_s,\omega_t,\omega_t)-W^\sigma_{npq}({\bf k};\omega_t+\omega_s,\omega_t,\omega_r).
	\end{eqnarray} 
	\noindent
	The operator $\mathcal{P}$ represents all possible intrinsic permutations among frequencies $ \omega_r ,\omega_s ,\omega_t$. That implies no permutation when all three frequencies are equal, three permutations when only one frequency is different as $(\omega_1,\omega_2,\omega_2)$, $(\omega_2,\omega_1,\omega_2)$, and $(\omega_2,\omega_2,\omega_1)$, and six permutations otherwise as $(\omega_1,\omega_2,\omega_3)$, $(\omega_1,\omega_3,\omega_2)$, $(\omega_2,\omega_1,\omega_3)$, $(\omega_2,\omega_3,\omega_1)$, $(\omega_1,\omega_2,\omega_3)$, and $(\omega_1,\omega_3,\omega_2)$.
	
	The functions $W^\sigma_{mnp}({\bf k};w_1,w_2,w_3)$ and $U_{pq}^\sigma({\bf k};w)$ are given as
	\begin{eqnarray}
	W^\sigma_{mnp}({\bf k};w_1,w_2,w_3)&=&\frac{U_{mn}^\sigma({\bf k};w_2)-U_{np}^\sigma({\bf k};w_3)}{\hbar w_1+E_{mp}^\sigma({\bf k})},\\ \nonumber
	U_{pq}^\sigma({\bf k};w)&=&\frac{f^\sigma_p({\bf k})-f^\sigma_q({\bf k})}{\hbar w+E_{pq}^\sigma({\bf k})}.
	\end{eqnarray}
	For the diamagnetic contribution, we have
	\begin{eqnarray}
	\Delta_{\mu\nu\zeta\eta}^{\sigma}({\bf k}; \omega_r ,\omega_s ,\omega_t )&&=\mathcal{P}\sum_{mnpq}[A_{mnpq}^{\mu\nu\zeta\eta\sigma}({\bf k}; \omega_r ,\omega_s ,\omega_t )-\\ \nonumber
	&& B_{mnpq}^{\mu\nu\zeta\eta\sigma}({\bf k}; \omega_r ,\omega_s ,\omega_t )-C_{mnpq}^{\mu\nu\zeta\eta\sigma}({\bf k}; \omega_r ,\omega_s ,\omega_t )],\\ 
	A_{mnpq}^{\mu\nu\zeta\eta\sigma}({\bf k}; \omega_r ,\omega_s ,\omega_t )&&=U_{mn}^\sigma({\bf k};\omega_r+\omega_s)g^{\mu\nu\sigma}_{mn}({\bf k}) g^{\zeta\eta\sigma}_{nm}({\bf k}),\\ 
	B_{mnpq}^{\mu\nu\zeta\eta\sigma}({\bf k}; \omega_r ,\omega_s ,\omega_t )&&=j_{pm}^{\mu\sigma}({\bf k})j_{nm}^{\nu\sigma}({\bf k})g_{mp}^{\zeta\eta\sigma}({\bf k})\\ \nonumber
	&&W^\sigma_{mnp}({\bf k};\omega_r+\omega_s,\omega_t,\omega_t),\\ 
	C_{mnpq}^{\mu\nu\zeta\eta\sigma}({\bf k}; \omega_r ,\omega_s ,\omega_t )&&=j_{nm}^{\eta\sigma}({\bf k})j_{np}^{\eta\sigma}({\bf k})g_{pm}^{\mu\nu\sigma}({\bf k})\\ \nonumber
	&& W^\sigma_{mnp}({\bf k};\omega_\Sigma,\omega_r+\omega_s,\omega_t).
	\end{eqnarray}
\end{widetext}
\noindent
The matrix element $g_{pm}^{\mu\nu\sigma}({\bf k})$ is obtained as
\begin{eqnarray}
g_{mn}^{\mu\nu\sigma}({\bf k})&=&\braket{\psi_m^\sigma({\bf k})|\hat{g}_{\mu\nu}^{\sigma}({\bf k})|\psi_n^\sigma({\bf k})},\\ \nonumber
\hat{g}_{\mu\nu}^{\sigma}({\bf k})&=&-\frac{q^2}{\hbar^2}\frac{\partial^2\mathbb{H({\bf k};\sigma)}}{\partial k_{\mu}\partial k_{\nu}}.
\end{eqnarray}

In all the above relationships, a small positive imaginary part is normally added to all three frequencies, in order to preserve causality as well as to avoid numerical overflow at resonances. Needless to say, coding these relations are not straightforward and needs extra care. The double integration over the reciprocal lattice could be done by evaluation of the integrand over a discrete triangular grid and multiplying by element sizes. Employing adaptive numerical integration methods is not practical because of the extremely large numerical burden. 

\section{Imaginary Squeezing}\label{AppC}

If the third-order nonlinear susceptibility $\chi^{(3)}$ is imaginary, then the two-photon absorption is dominant over the Kerr effect. While the Kerr effect could in principle produce unlimited squeezing, two-photon absorption may cause a limited squeezing of shot noise. This effect has been noticed by a number of authors in the past \cite{Fano1,Fano2,Fano3,Fano4,Fano5,Fano6,Fano7}. However, in all these works the temporal evolution of squeeze parameter is considered, in which squeezing generally increases with propagation, reaching an ultimate value in the limit of infinite propagation in a two-photon absorbing medium. That type of analysis is useful for nonlinearly lossy long fibers. 

Here, for the case of a ring resonator with constant propagation length, it is the intensity, or equivalently, the cavity photon number $N$,  of the input beam which could be varied. In what follows, we show that squeezing generally increases with the input power, first within the approximation of negligible loss over propagation, that is constant $\zeta$.

\subsection{Non$-$rotated Squeezing}

\begin{figure}[ht!]
	\centering
	\includegraphics[width=2.2in]{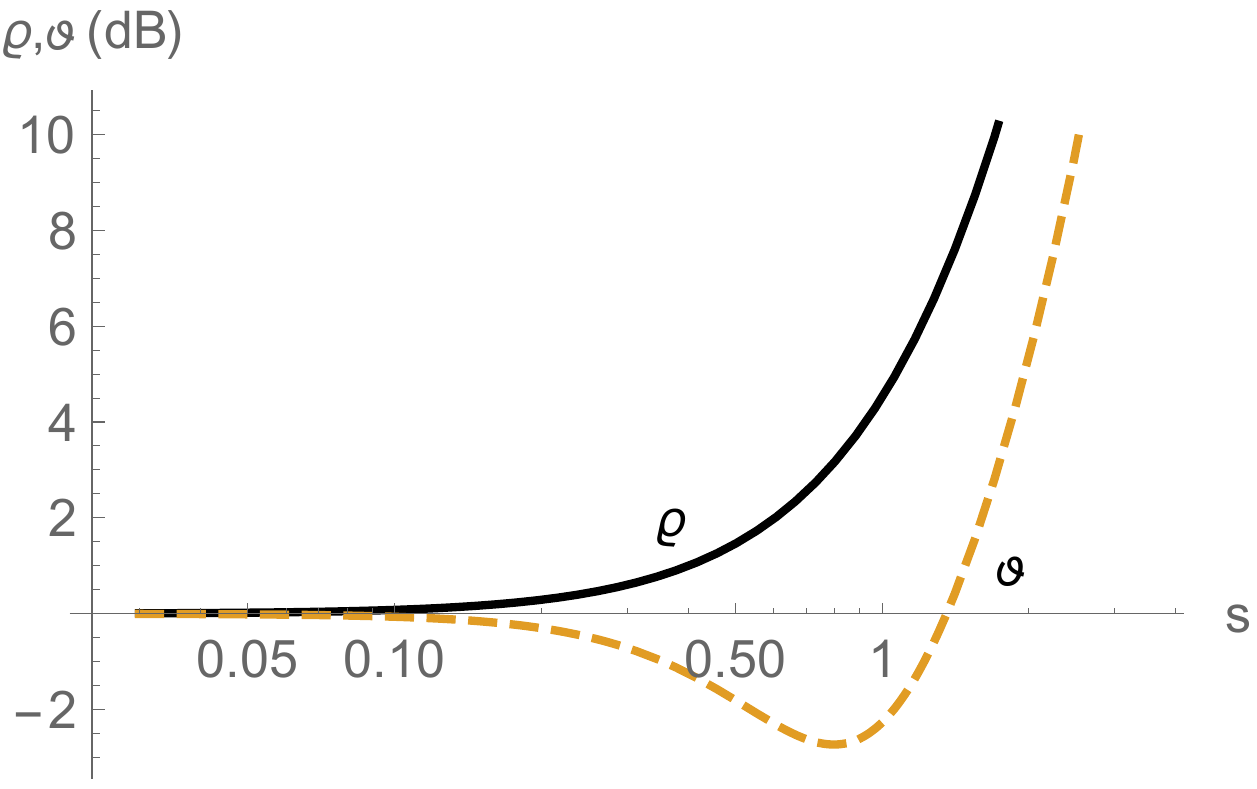}
	\centering
	\caption{Variation of squeezing amplitudes versus absolute value of purely imaginary squeeze parameter. Both quadratures desqueeze for large $s$.\label{FigSq}}
\end{figure}

The wavefunction corresponding to the squeezed coherent state $\ket{\alpha,\zeta}$ with $\alpha$ being the complex coherent state number, generated by an interaction of the type (\ref{eq7a}) or equivalently produced from a coherent state such as $\ket{\alpha}$ by application of the operator $\hat{S}(\zeta)$ (\ref{sq1}), is given by \cite{Nieto1,Nieto2,Miri}
\begin{equation}
\braket{x|\alpha,\zeta}=\frac{e^{i(x-x_0)p_0}}{\pi^\frac{1}{4}C}\exp\left[-\left(\frac{1}{2gC^2}-ih\right)(x-x_0)^2\right],
\end{equation}
\noindent
in which $x$ is in the units of zero-point fluctuations $x_{\rm zp}$, $x_0=\Re[\alpha]$, $p_0=\Im[\alpha]$, and 
\begin{eqnarray}
g&=&\cosh{|\zeta|}+\frac{\Re[\zeta]}{|\zeta|}\sinh{|\zeta|},\\ \nonumber
h&=&\frac{\Im[\zeta]\sinh{|\zeta|}}{2|\zeta|e^{|\zeta|}},\\ \nonumber
C&=&\sqrt{g(1+2ih)}.
\end{eqnarray}
\noindent
It is straightforward to verify via Fourier transformation that the momentum representation would be given by
\begin{equation}
\braket{p|\alpha,\zeta}=\frac{e^{ix_0(p-p_0)}}{\pi^\frac{1}{4}\sqrt{\frac{1}{g}-2ihC}}\exp\left[-\frac{(p-p_0)^2}{2\left(\frac{1}{gC^2}-2ih\right)}\right].
\end{equation}
\noindent
Since we have assumed that $\zeta=is$, we have
\begin{eqnarray}
g&=&\cosh{s},\\ \nonumber
h&=&\frac{\sinh{s}}{2e^s}.
\end{eqnarray}
Therefore, the probablity distribution in position is given by 
\begin{equation}
|\braket{x|\alpha,\zeta}|^2=\frac{1}{\sqrt{\pi}|C|^2}\exp\left\{-\Re\left[\frac{1}{gC^2}\right](x-x_0)^2\right\},
\end{equation}
where we have noticed that $h$ is real-valued. Similarly, for the probablity distribution in the momentum representation we obtain 
\begin{equation}
|\braket{p|\alpha,\zeta}|^2=\frac{\exp\left\{-\Re\left[\left(\frac{1}{gC^2}-2ih\right)^{-1}\right](p-p_0)^2\right\}}{\sqrt{\pi}|\frac{1}{g}-2ihC|},
\end{equation}
Comparing to the Gaussian distribution of a simple coherent state \cite{10}, we may deduce the noise squeezing ratio $\varrho$ in position and $\varphi$ in momentum, also known as the Fano factor \cite{Fano7}, respectively by solving the equations 
\begin{eqnarray}
\varrho^{-2}&=&\Re\left[\frac{1}{gC^2}\right]=\frac{1}{g^2(1+4h^2)}, \\ \nonumber
\varphi^{-2}&=&\frac{\Re\left[\frac{1}{gC^2}\right]}{|\frac{1}{gC^2}-2ih|^2}=\frac{g^2}{(1+4h^2g^2)^2+4h^2g^4}.
\end{eqnarray}
\noindent
Interestingly, for a purely real $\zeta$ with $\Im[\zeta]=0$, we have $C=\sqrt{g}$, $h=0$, and this expression takes the simple solution $\varrho=\exp(\zeta)$ and $\varphi=\exp(-\zeta)$. This corresponds to a minimum uncertainty squeezed packet  \begin{equation}
\Delta\varphi\Delta\varrho=\frac{1}{2},
\end{equation} 
\noindent
with $\Delta\varphi=\varphi/\sqrt{2}$ and $\Delta\varrho=\varrho/\sqrt{2}$. But for the present case where $\Re[\zeta]=0$, after some alegbraic manipulations it gives the solutions
\begin{eqnarray}
\label{Squeeze}
\varrho&=&\cosh(s)\sqrt{1+e^{-2s}\sinh^2(s)},\\ \nonumber
\varphi&=&\sqrt{\frac{\sinh^2(2s)}{4e^{2s}}+{\rm sech}^2(s)\left[1+\frac{\sinh^2(2s)}{4e^{2s}}\right]^2}.
\end{eqnarray}
It is easy to observe that for large values of $s$, we get $\varrho\approx2\varphi\approx(\sqrt{5}/4)e^{s}$, which represents unlimited desqueezing and shot noise increase. A plot of $\varrho$ and $\varphi$ versus $s$ is illustrated in Fig. \ref{FigSq}. The momentum quadrature exhibits a minimum of $\varphi_{\min}=-2.73{\rm dB}$ at $s_{\min}=0.797$

\subsection{$\frac{\pi}{4}-$rotated Squeezing}

While non-rotated quadratures both appear to be strongly desqueezed, the fact is that the true squeezing actually happens along the $\frac{\pi}{4}-$rotated quadratures. To illustrate this, we need to rigorously calculate the Wigner function of the squeezed vacuum first. This is given by
\begin{eqnarray}
W_{\hat{S}(\zeta)\ket{0}}(\theta)&=&\frac{1}{\pi^2}\iint \exp(\theta\beta^*-\theta^*\beta) \\ \nonumber
&&\exp\left[-\frac{1}{2}|\beta\mu+\beta^*\nu|^2\right]d^2\beta,
\end{eqnarray}
\noindent
where $\theta=x+ip$, $\beta=a+ib$, $\mu=\cosh|\zeta|$, and $\nu=\exp(i\angle\zeta)\sinh|\zeta|$. Using the fact that $\zeta=is$, and some straightforward but significant algebra, we get the Wigner distribution of purely-imaginary squeezed vacuum as 
\begin{eqnarray}
W_{\hat{S}(\zeta)\ket{0}}(x,p)&=&\frac{2}{\pi}\exp\left[-\frac{2p^2}{\cosh(2s)}\right] \times \\ \nonumber
&& \exp\left\{-2\cosh(2s)\left[x+\tanh(2s)p\right]^2\right\}.
\end{eqnarray}

\begin{figure}[ht!]
	\centering
	\includegraphics[width=2.2in]{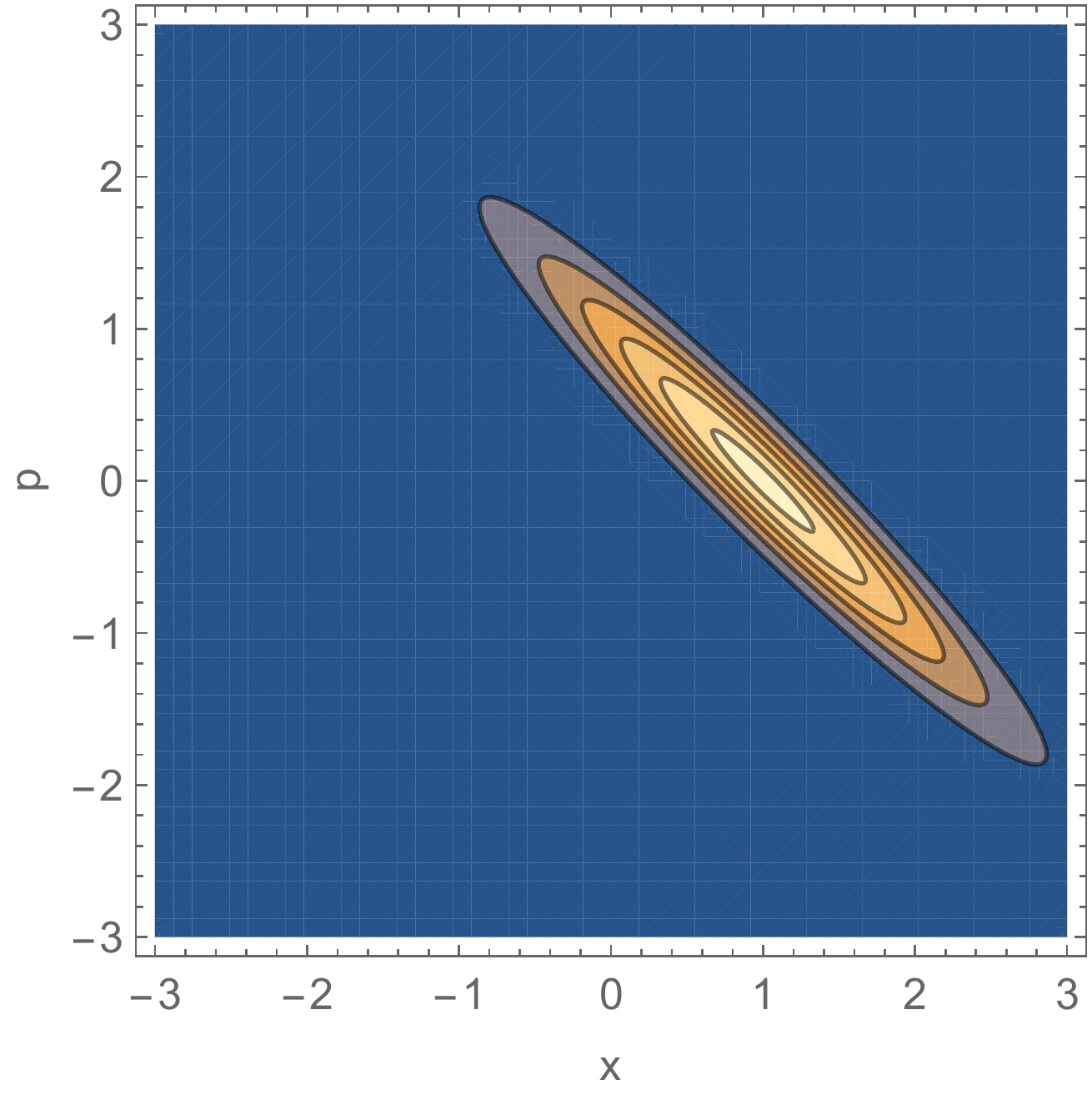}
	\centering
	\caption{Wigner distribution of the squeezed coherent state with $\alpha=1$ and $\zeta=i$. Both non-rotated orthogonal quadratures $(x,p)$ appear to be desqueezed almost to the same amount. \label{Wigner}}
\end{figure}

\begin{figure}[ht!]
	\centering
	\includegraphics[width=2.2in]{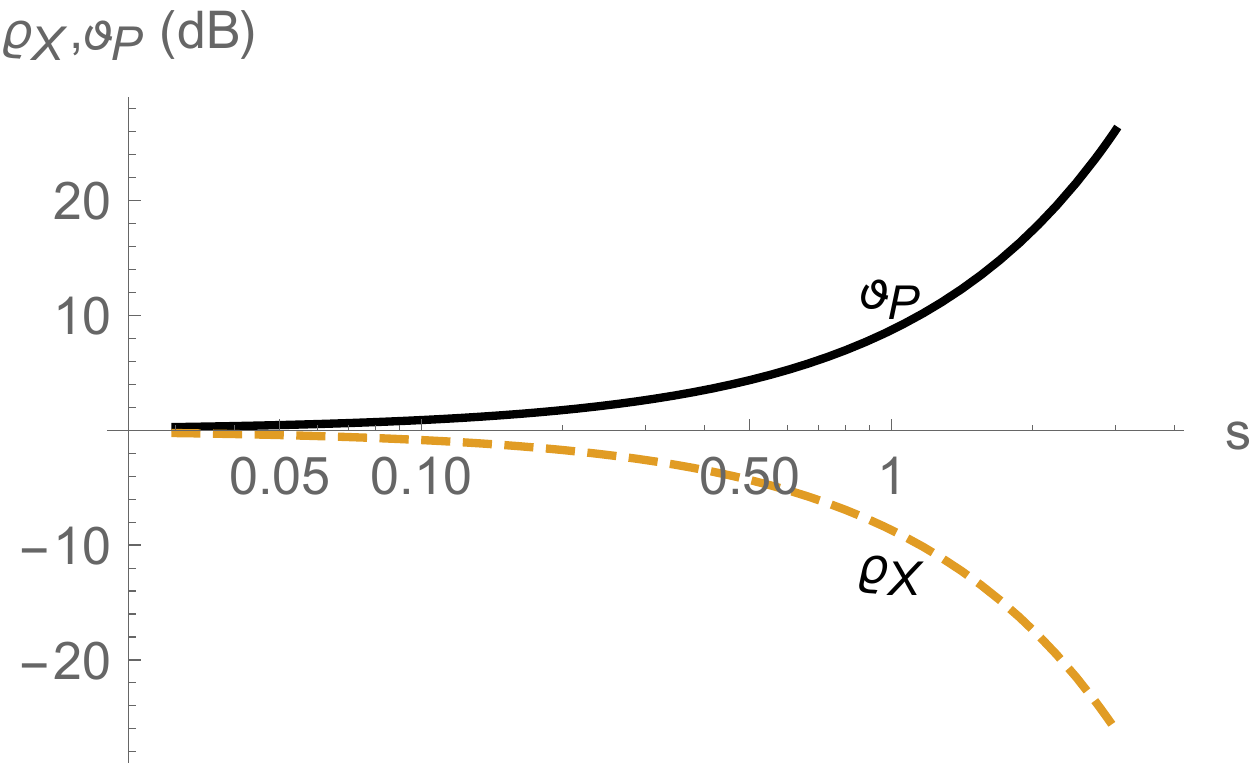}
	\centering
	\caption{Variation of squeezing amplitudes in the $\frac{\pi}{4}-$rotated system versus absolute value of purely imaginary squeeze parameter. One rotated quadrature is squeezed and the other is strongly desqueezed. \label{45rotated}}
\end{figure}
\noindent
The probablity distributions obtained as \cite{10}
\begin{eqnarray}
|\braket{x|\alpha,\zeta}|^2&=&\int W_{\hat{D}(\alpha)\hat{S}(\zeta)\ket{0}}(x,p)dp, \\ \nonumber
|\braket{p|\alpha,\zeta}|^2&=&\int W_{\hat{D}(\alpha)\hat{S}(\zeta)\ket{0}}(x,p)dx,
\end{eqnarray}
from the above Wigner function both appear to be desqueezed almost equally, as discussed in the above and shown in Fig. \ref{FigSq} for large $s$. Here, the displacement operator $\hat{D}(\alpha)$ which produces a coherent state simply transforms $x\rightarrow x-\Re[\alpha]$ and $p\rightarrow p-\Im[\alpha]$. In the the time-dependent case, we only need to replace $\alpha=A\exp(i\omega t)$ where $A$ is the amplitude of the coherent source and $\omega$ is the angular frequency of light. The typical resulting Wigner function for the squeezed coherent state is illustrated in Fig. \ref{Wigner}.

Now, we can define the $\frac{\pi}{4}-$rotated coordinates as
\begin{eqnarray}
x&=&\frac{X+P}{\sqrt{2}}, \\ \nonumber
p&=&\frac{X-P}{\sqrt{2}}.
\end{eqnarray}
\noindent
In this new system of coordinates, we obtain the Wigner distribution of the squeezed state $\ket{\alpha,\zeta}=\hat{D}(\alpha)\hat{S}(is)\ket{0}$ as
\begin{eqnarray}
W_{\ket{\alpha,\zeta}}(X,P)&=&\frac{2}{\pi}\exp\left[-2e^{2s}(X-\Re[\alpha])^2\right] \\ \nonumber
&& \times\exp\left[-2e^{-2s}(P-\Im[\alpha])^2\right],
\end{eqnarray}
\noindent
with the respective squeeze ratios
\begin{eqnarray}
\varrho_X&=&\sqrt{2}\Delta\varrho=e^{-s}, \\ \nonumber
\varphi_P&=&\sqrt{2}\Delta\varphi=e^{s},
\end{eqnarray}
\noindent
plotted in Fig. \ref{45rotated}, and giving the uncertainty product 
\begin{equation}
\Delta\varrho\Delta\varphi=\frac{1}{2},
\end{equation}
\noindent
which refers to the minimum-uncertainty packet. 

\section{Units of Surface Quantities}\label{AppD}

While the linear bulk susceptibility $\chi^{(1)}$ should be dimensionless, the surface linear susceptibility $\chi^{(1){\rm 2D}}$ is not, actually having the dimension of length. But it is preferable to write the dimension as ${\rm m}\Box$ or ${\rm\AA}\Box$ where the redundant dimensionless square notation $\Box$ emphasizes a 2D surface quantity. Similarly, the dimension of nonlinear susceptibility $\chi^{(3){\rm 2D}}$ would be preferably ${\rm m^3V^{-2}}\Box$ instead of ${\rm m^3V^{-2}}$. It follows then, that the preferable dimension of linear and nonlinear sheet conductivity $\sigma^{(3){\rm 2D}}$ must be respectively $\Omega^{-1}\Box$ and $\Omega^{-1}{\rm m^{2}V^{-2}}\Box$.

\vspace{5mm}

\bibliography{X3Ref}
\end{document}